# Entropy and Adjoint Methods

Carlos Lozano
*Computational Aerodynamics Group*
*National Institute for Aerospace Technology (INTA)*

**Abstract**

Aerodynamic drag can be partially approximated by the entropy flux across fluid domain boundaries with a formula due to Oswatitsch. In this paper, we build the adjoint solution that corresponds to this representation of the drag and investigate its relation to the entropy variables, which are linked to the integrated residual of the entropy transport equation. For inviscid isentropic flows, the resulting adjoint variables are identical to the entropy variables, an observation originally due to Fidkowski and Roe, while for non-isentropic flows there is a significant difference that is explicitly demonstrated with analytic solutions in the shocked quasi-1D case. Both approaches are also investigated for viscous and inviscid flows in two and three dimensions, where the adjoint equations and boundary conditions are derived. The application of both approaches to mesh adaptation is investigated, with especial emphasis on inviscid flows with shocks.

## 1 Introduction

In computational aerodynamics, adjoint methods can be applied to aerodynamic design [1], which was their original motivation, but also to error estimation and goal-oriented mesh adaptation [2, 3]. Numerous adjoint-based mesh adaptation algorithms targeting typical cost functions such as lift or drag have been reported (see [4] for a recent example), but other targets are possible. In this paper, the focus is on entropy generation, which (particularly in the case of spurious production) has been a significant ingredient to judge the accuracy of numerical solutions since the early days of CFD [5].

Many systems of conservation laws are equipped with an entropy function, which is a convex function of the system state variables that is conserved for smooth solutions but can exhibit discontinuities for non-smooth (weak) solutions. The sign of the jump at discontinuities of the solution serves as a unicity criterion singling out the physically realizable solution (entropy condition). Besides, entropy functions play an important role in the stability theory of PDEs. The gradient of the entropy function with respect to the state variables defines the entropy variables, which symmetrize the governing equations [6] [7] and are an important ingredient in the construction of entropy conservative/stable discretizations [8].

Aside from these formal developments, practical applications of entropy and entropy-based functionals in numerical simulations abound (see [9, 10] for reviews). In [11], a shape design procedure for turbomachinery applications based on minimization of entropy losses was proposed. In [12, 13, 14, 15, 16, 17, 18, 19], different forms of the residual of the entropy transport equation have been used as indicators for grid adaptation.

Reference [16] also contains the interesting observation that, for the Euler equations, the entropy variables are actually adjoint variables associated to the integrated entropy residual and, thus, with spurious entropy production. The adjoint approach based on the entropy variables (or entropy adjoint approach) can also deal with viscous flows, where it is also associated to the integrated entropy residual, although in this case the entropy variables do not obey an adjoint equation.

Another significant observation was made in [17] concerning the connection of the entropy adjoint approach to aerodynamic drag. Viscous and wave drag can be approximated, with a formula due to Oswatitsch [20], by the net entropy flux across inlet/exit and/or far-field boundaries, which is also directly related to loss in turbomachines [21]. For inviscid flows without shocks, this boundary entropy balance is equivalent to the integrated entropy residual, so entropy variables are adjoint to the Oswatitsch representation of aerodynamic drag. Unfortunately, this is actually spurious drag, as inviscid smooth flows have zero drag analytically. When the flow contains shocks, physical drag is proportional to the entropy created at shocks and no longer vanishes. On the other hand, entropy variables are still adjoint to the integrated entropy residual, which, as will be shown below (see also [22]), can be written as the net entropy balance across all domain boundaries and flow discontinuities (including shock loci) or, equivalently, as the balance between the entropy entering and leaving the domain and the entropy generated within the domain. Using the entropy variables thus poses the practical issue that discretized entropy sensors based on the entropy residual contain the (physical) entropy produced at the shock, which may cause trouble in mesh adaptation as discussed for example in [19]. Several ways to address this issue, including the comparison with the Oswatitsch adjoint, will be analyzed below.

The paper is organized as follows. Section 2 reviews the connection between entropy flux and aerodynamic drag. Section 3 deals with the formulation of the entropy and Oswatitsch adjoints for quasi-one-dimensional (quasi-1D) flows with shocks, including a proposal for local error indicator as well as numerical testing. Sections 4 and 5 deal with two-dimensional (2D) and three-dimensional (3D) inviscid shocked flows, respectively, while section 6 deals with 2D (laminar) viscous flows. Finally, Section 7 contains a brief summary and discussion of the results.

## 2 Drag and entropy flux

Aerodynamic drag is the force exerted by a moving fluid on a solid body along the direction $\vec{d}$ of the body's motion. Focusing on the configuration depicted in Fig. 1, consisting on a body with a solid wall surface $S_w$ immersed in a fluid domain with external (far-field) boundary $S_\infty$, drag can be computed as the integral of the fluid's momentum flux over the body surface. For inviscid flows, it reduces to

$$D_{near} = \int_{S_w} p\vec{d}\cdot\hat{n}\,dS \qquad (1)$$

where $\hat{n}$ is the outward-pointing unit normal vector and $p$ is the pressure. Since the momentum flux is conserved, the same result can be obtained integrating the momentum flux over the fluid domain's outer boundary $S_\infty$

$$D_{ff} = -\int_{S_\infty}\left(p\vec{d}\cdot\hat{n} + \rho(\vec{d}\cdot\vec{u})(\vec{u}\cdot\hat{n})\right)dS$$

where $\rho$ and $\vec{u}$ are the fluid's density and velocity, respectively. The above drag integral can be related to the entropy flux across the far-field boundary as follows. Under suitable assumptions [17] [23], the part of fluid flow drag due to shocks or boundary layers can be computed by the following far-field integral (first derived by Oswatitsch [20]):

$$D_{osw} \approx \frac{\vec{u}_\infty\cdot\vec{d}}{\gamma RM_\infty^2}\int_{S_\infty}\rho\Delta s\vec{u}\cdot\hat{n}\,dS = -\frac{\vec{u}_\infty\cdot\vec{d}}{\gamma M_\infty^2}\int_{S_\infty}\vec{\Phi}\cdot\hat{n}\,dS \qquad (2)$$

where $s = \frac{R}{\gamma-1}\log(p/\rho^\gamma)$ is the entropy, $\vec{\Phi} = -\rho s\vec{u}/R$ the entropy flux, $R$ the gas constant, $\vec{u}_\infty$ the velocity of the incoming flow (the free-stream velocity) and $M_\infty$ the free-stream Mach number.

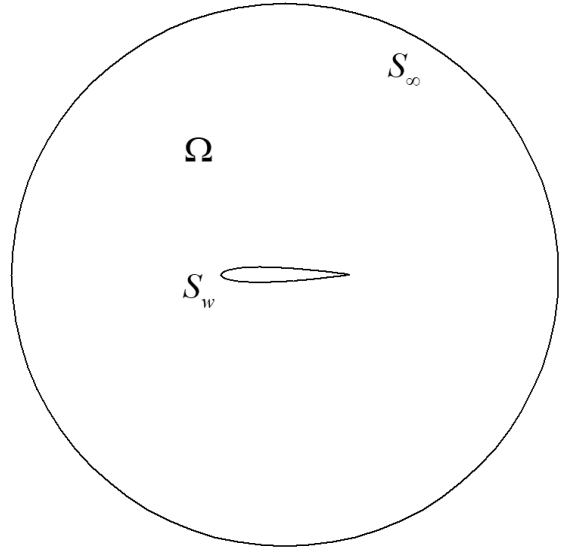

**Fig. 1 Sketch of a typical 2D flow configuration.**

In practice, other approaches for drag estimation that are indirectly based on the far-field entropy balance are used [24] besides conventional wall integration. These approaches allow for a clear separation of the different physical sources of drag and can be exploited in mesh adaptation [25].

## 3  Quasi-one-dimensional Euler equations

We shall begin our analysis with the (steady) quasi-1D Euler equations. On a duct of cross section $A(x)$, the equations read

$$R(U,A) = \frac{d}{dx}(AF) - \frac{dA}{dx}P = 0. \tag{3}$$

In the above equation, $U$ is the vector of conservation variables, $F$ are the mass, momentum and energy fluxes, and $P$ is the source term

$$U = \begin{pmatrix} \rho \\ \rho u \\ \rho E \end{pmatrix}, \quad F = \begin{pmatrix} \rho u \\ \rho u^2 + p \\ \rho u H \end{pmatrix}, \quad P = \begin{pmatrix} 0 \\ p \\ 0 \end{pmatrix} \tag{4}$$

where $\rho$ is the density, $u$ is the velocity and $p$, $E$ and $H$ are the pressure, energy and enthalpy, respectively. Also

$$p = (\gamma - 1)\rho\left(E - \frac{1}{2}u^2\right), \quad H = E + \frac{p}{\rho} = \frac{\gamma}{\gamma - 1}\frac{p}{\rho} + \frac{1}{2}u^2 \tag{5}$$

where $\gamma = c_p / c_v$ is the ratio of specific heats.

In the absence of shock waves, the entropy $s = c_v \log(p/\rho^\gamma)$ is constant and obeys the following advection equation

$$\frac{d(\rho u A s)}{dx} = 0 \tag{6}$$

or, equivalently, $ds/dx = 0$, since $d(\rho u A)/dx = 0$ as follows from the governing equations (3). Suppose that the duct extends along the interval $-1 \leq x \leq 1$. Integrating (6) along the duct yields

$$0 = \int_{-1}^{+1} \frac{d}{dx}(\rho u A s)\, dx = [\rho u A s]_{-1}^{+1} \tag{7}$$

where the rightmost expression gives the net entropy balance across the inlet/exit boundaries (which vanishes in isentropic –shock-free– cases).

If the solution contains a steady shock at $x_s$, (6) holds on either side of the shock, in such a way that integrating (6) along the duct yields the following relation

$$0 = \int_{-1}^{x_s^-} \frac{d}{dx}(\rho u A s)\, dx + \int_{x_s^+}^{+1} \frac{d}{dx}(\rho u A s)\, dx = [\rho u A s]_{-1}^{+1} - [\rho u A s]_{x_s^-}^{x_s^+} \tag{8}$$

where the rightmost expression gives the net entropy balance across the domain boundaries including the shock.

### 3.1 Entropy variables, the Oswatitsch adjoint and the entropy adjoint approach

Many entropy functions can be built for system (3). A particularly simple one is

$$\eta = -\rho s / R = -\frac{\rho}{\gamma - 1}\log(p/\rho^\gamma) \tag{9}$$

which is proportional to the physical entropy. For steady flows, $\eta$ obeys the following conservation law

$$\frac{d(A\Phi)}{dx} = 0 \tag{10}$$

where the entropy flux is

$$\Phi = u\eta = -\rho u s / R \tag{11}$$

The derivatives of $\eta$ with respect to the conservative variables,

$$v^T = \frac{\partial \eta}{\partial U} = \left( \frac{\gamma}{\gamma-1} - \frac{s}{R} - \frac{\rho u^2}{2p}, \frac{\rho u}{p}, -\frac{\rho}{p} \right) \tag{12}$$

are the *entropy variables*, and they have several interesting properties. Most notably, they obey the compatibility condition

$$v^T \frac{\partial F}{\partial U} = \frac{\partial \Phi}{\partial U}, \tag{13}$$

and they symmetrize the system (3), as both $U_v$ and $F_v$ are symmetric. Likewise, multiplying (3) by v and using (13), the entropy conservation law (10) is recovered

$$v^T R(U, A) = \frac{d}{dx}(A\Phi) = 0 \tag{14}$$

Integrating (14) along the duct yields the following relation

$$0 = \int_{-1}^{+1} v^T R(U, A) dx = \int_{-1}^{+1} \frac{d}{dx}(A\Phi) dx = [A\Phi]_{-1}^{+1} \tag{15}$$

which relates the entropy-weighted residual with the isentropic entropy balance (7).

The symmetrization properties of the entropy variables [7] imply that both $U$ and $F$ can be expressed as gradients of scalar functions, $U(v) = \nabla_v \theta$, $F(U(v)) = \nabla_v \Theta$, and thus the entropy / entropy flux pair can be expressed as

$$\eta = v^T U - \theta, \quad \Phi = v^T F - \Theta. \tag{16}$$

Note that for the above choices, the entropy potentials are $\theta = \rho$ and $\Theta = \rho u$, respectively.

Further, note that, by writing the flow equations in terms of the entropy variables and using their symmetry properties, it follows that the entropy variables satisfy an adjoint-like equation

$$AF_U^T \frac{dv}{dx} + \frac{dA}{dx} P_U^T v = 0 \tag{17}$$

Eq. (15) and (17) are the basis of the observation in [16] that the entropy variables are adjoint variables relative to the output

$$J = [A\Phi]_{-1}^{+1} \tag{18}$$

measuring the entropy flux across the fluid domain outer boundaries. Indeed, the adjoint state $\psi = (\psi_0, \psi_1, \psi_2)^T$ corresponding to (18) can be derived from the analysis of the linear perturbations to the Lagrangian

$$J = [A\Phi]_{-1}^{+1} - \int_{-1}^{1} \psi^T R(U, A) dx \tag{19}$$

After linearization with respect to *A* and *U*, integration by parts and rearrangement, we get,

$$\delta J = [\delta A \Phi]_{-1}^{+1} + [A\Phi_U \delta U]_{-1}^{+1} - \int_{-1}^{1} \psi^T \delta R(U,A)dx =$$
$$[\delta A \Phi]_{-1}^{+1} + \int_{-1}^{1} f^T \psi dx - \int_{-1}^{1} \delta U^T L^* \psi dx + \left[ A(\Phi_U - \psi^T F_U)\delta U \right]_{-1}^{+1}$$
(20)

where

$$f = \frac{d\delta A}{dx} P - \frac{d}{dx}(\delta A F)$$
$$L^* \psi = -AF_U^T \frac{d\psi}{dx} - \frac{dA}{dx} P_U^T \psi$$
(21)

To make (26) independent of $\delta U$, $\psi$ must obey the adjoint equation

$$-AF_U^T \frac{d\psi}{dx} - \frac{dA}{dx} P_U^T \psi = 0,$$
(22)

with the inlet/exit boundary conditions

$$(\Phi_U - \psi^T F_U)\delta U \big|_{x=\pm 1} = (v - \psi)^T F_U \delta U \big|_{x=\pm 1} = 0$$
(23)

from where it follows that $\psi = v$, i.e., the entropy variables are adjoint to the entropy flux balance across the inlet/exit boundaries, a relation which also holds (with obvious modifications) in 2D and 3D. Hence, the entropy variables are adjoint, via the Oswatitsch formula, to the far-field drag integral addressed in [17]. Although drag by itself is of no interest for subsonic inviscid solutions (since its exact value is zero), v can still be used to build an adaptation indicator that increases the accuracy of the solution by targeting the spurious contributions to the far-field drag.

When the flow contains a shock, the above derivation needs to be corrected and, in fact, the entropy variables are now adjoint variables to the net entropy balance across the domain boundaries (including the shock loci). This can be seen as follows. If the solution contains a steady shock at $x_s$, the flow equations $R(U,A) = 0$ hold on either side, while at the shock the solution obeys the Rankine-Hugoniot jump condition $[F]_{x_s^-}^{x_s^+} = 0$. Eq. (14) also remains valid on either side of the shock, in such a way that integrating along the duct in (14) yields the following relation

$$0 = \int_{-1}^{x_s^-} v^T R(U,A)dx + \int_{x_s^+}^{+1} v^T R(U,A)dx =$$
$$\int_{-1}^{x_s^-} \frac{d(A\Phi)}{dx} dx + \int_{x_s^+}^{+1} \frac{d(A\Phi)}{dx} dx = [A\Phi]_{-1}^{+1} - [A\Phi]_{x_s^-}^{x_s^+}$$
(24)

which relates the entropy-weighted residual with the entropy balance (8). In order to prove that the entropy variables are indeed adjoint to the output (24), we analyze the linear perturbations to the Lagrangian

$$J = [A\Phi]_{-1}^{+1} - [A\Phi]_{x_s^-}^{x_s^+} - \int_{-1}^{x_s^-} \psi^T R(U,A)dx - \int_{x_s^+}^{+1} \psi^T R(U,A)dx - A_s \varphi^T [F]_{x_s^-}^{x_s^+}$$
(25)

(with $A_s = A(x_s)$), where we have introduced the adjoint states $\psi, \varphi$ to enforce the flow and shock equations. After linearization, integration by parts and rearrangement, we obtain

$$\delta J = \left[\delta A \Phi\right]_{-1}^{+1} - \delta A_s \left[\Phi\right]_{x_s^-}^{x_s^+} - \delta x_s \left[\frac{d(A\Phi)}{dx}\right]_{x_s^-}^{x_s^+} - A_s \varphi^T \left[\frac{dF}{dx}\right]_{x_s^-}^{x_s^+} \delta x_s$$
$$+ \int_{-1}^{x_s^-} \psi^T f \, dx + \int_{x_s^+}^{+1} \psi^T f \, dx - \int_{-1}^{x_s^-} \delta U^T L^* \psi \, dx - \int_{x_s^+}^{1} \delta U^T L^* \psi \, dx \quad (26)$$
$$+ \left[A_s (\psi - v - \varphi)^T F_U \delta U\right]_{x_s^-}^{x_s^+} + \left[A(v - \psi)^T F_U \delta U\right]_{-1}^{+1}$$

with $f$ and $L^* \psi$ as in Eq. (21). Since $\left[d(A\Phi)/dx\right]_{x_s^-}^{x_s^+} = 0$ because of eq. (10), removing the dependence of (26) on $\delta x_s$ requires that $\varphi = 0$. To eliminate the dependence on $\delta U$, $\psi$ must obey the adjoint equation (22) in $x \in [-1, x_s) \cup (x_s, +1]$ with the inlet/exit boundary conditions (23), as well as the internal shock conditions

$$\left[(\psi - v)^T F_U \delta U\right]_{x_s^-}^{x_s^+} = 0 \quad (27)$$

Taking (17) into account, the above equations require that $\psi = v$ throughout, thus confirming that the entropy variables are indeed the adjoint state associated to $J = \left[A\Phi\right]_{-1}^{+1} - \left[A\Phi\right]_{x_s^-}^{x_s^+}$, which is obviously different from the inlet/exit entropy balance $J = \left[A\Phi\right]_{-1}^{+1}$. The adjoint state relative to $J = \left[A\Phi\right]_{-1}^{+1}$ in the shocked case can be built following the approach outlined above. The resulting Oswatitsch adjoint $\psi$ obeys the adjoint equation (22) on either side of the shock with the inlet/exit boundary conditions (23). Additionally, $\psi$ must satisfy the following shock conditions

$$\psi^T F_U \delta U \big|_{x_s^+} - \psi^T F_U \delta U \big|_{x_s^-} = 0 \quad (28)$$

which requires that the adjoint variables be continuous at the shock with zero derivative and obey the internal boundary condition $\psi_1(x_s) = 0$ [26]. The above equations are incompatible with $\psi$ being the entropy variables, which are discontinuous at the shock (see Fig. 3).

### 3.2 Discrete analysis

As with most existing a posteriori error estimators, adjoint-based methods require either the local residual error $R(U_h, A)$, i.e., the continuous state equation $R(\cdot, A)$ evaluated at (a suitable continuous reconstruction of) the discrete solution $U_h$, or the truncation error $R_h(U_{exact})$, obtained by substituting the analytic solution $U_{exact}$ (or a suitable approximation thereof) into the discrete operator. The adjoint error correction term is a product of the approximate adjoint solution and the residual/truncation error. The latter may be approximated by the integral of the analytic flux about each control volume [27], by the nonlinear residual evaluated on a continuous interpolation of the discrete solution [28], by a discrete residual evaluated at a higher order of accuracy, or on a uniformly refined mesh [29], or on a coarser mesh via the τ-estimation approach [30, 31, 32]; or even by the discrete artificial dissipation residual [33].

We have seen above that the integrated entropy balance equation residual $J$ is related to the analytic flow equations as

$$J = \int_{-1}^{+1} v(U)^T R(U, A) dx \tag{29}$$

which is clearly of the form described above. In order to build an error indicator based on the entropy variables, some discretized form of (29) must be constructed. In order to explore this idea, we shall consider the discretized version of the problem discussed in the preceeding section. Suppose then that (the time-varying version of) eq. (3) is approximated numerically on a grid over the range $i = 1, n$ by a semi-discrete finite-volume (FV) scheme

$$\Delta x_i A_i \frac{dU_{h,i}}{dt} + R_{h,i}(U_h) = 0 \tag{30}$$

where the residuals are $R_{h,i}(U_h) = A_{i+\frac{1}{2}} F^*_{i+\frac{1}{2}} - A_{i-\frac{1}{2}} F^*_{i-\frac{1}{2}} - (A_{i+\frac{1}{2}} - A_{i-\frac{1}{2}}) P_i$ and $F^*_{i+\frac{1}{2}}$ is the numerical flux function. Multiplying the LHS of (30) by the entropy variables $v^T_{h,i}$ and rearranging yields the semidiscrete numerical entropy scheme [8]

$$\Delta x_i A_i \frac{d\eta_{h,i}}{dt} + A_{i+\frac{1}{2}} \Phi^*_{i+\frac{1}{2}} - A_{i-\frac{1}{2}} \Phi^*_{i-\frac{1}{2}} = \Pi_i \tag{31}$$

where

$$\Phi^*_{i\pm\frac{1}{2}} = \overline{v}^T_{i\pm\frac{1}{2}} F^*_{i\pm\frac{1}{2}} - \overline{\Theta}_{i\pm\frac{1}{2}} \tag{32}$$

is the numerical entropy flux ($\Theta_i = \rho_i u_i$ is the entropy potential (16), and the overbar denotes averaging across the face, i.e., $\overline{\Theta}_{i+\frac{1}{2}} = \frac{1}{2}(\Theta_{i+1} + \Theta_i)$). Likewise,

$$\begin{aligned} \Pi_i &= \frac{1}{2}\left(\Pi_{i+\frac{1}{2}} + \Pi_{i-\frac{1}{2}}\right) \\ \Pi_{i\pm\frac{1}{2}} &= \Delta v^T_{i\pm\frac{1}{2}} F^*_{i\pm\frac{1}{2}} - \Delta \Theta_{i\pm\frac{1}{2}} \end{aligned} \tag{33}$$

(with $\Delta(\cdot)_{i+\frac{1}{2}} = (\cdot)_{i+1} - (\cdot)_i$ the jump across the face) is the entropy production. We note that numerical fluxes for which $\Pi_{i\pm\frac{1}{2}} = \Delta v^T_{i\pm\frac{1}{2}} F^*_{i\pm\frac{1}{2}} - \Delta \Theta_{i\pm\frac{1}{2}} = 0$ are entropy-conservative as proved by Tadmor [34].

From (30) and (31) we have

$$v^T_{h,i} R_{h,i} = A_{i+\frac{1}{2}} \Phi^*_{i+\frac{1}{2}} - A_{i-\frac{1}{2}} \Phi^*_{i-\frac{1}{2}} - \Pi_i \tag{34}$$

The weighted residual on the LHS has the conventional form of an adjoint-based error indicator. Unfortunately, for converged steady solutions the residuals are zero, and thus $v^T_{h,i} R_{h,i}$ is not a useful indicator. Assuming that the numerical flux can be written as a central flux plus dissipation, $F^*_{i+\frac{1}{2}} = \overline{F}_{i+\frac{1}{2}} - d_{i+\frac{1}{2}}$, the residual can be equally separated into

"convective" + "dissipation" parts, $R_{h,i} = R_{h,i}^{(c)} + R_{h,i}^{(diss)}$, such that the following indicator could be used $v_{h,i}^T R_{h,i}^{(c)}$. For converged steady solutions $R_{h,i} = 0$, so $v_{h,i}^T R_{h,i}^{(c)} = -v_{h,i}^T R_{h,i}^{(diss)}$, which is closely related to the dissipation-based adjoint error estimate introduced by Dwight [33] that we will use below. The only drawback is that, for shocked solutions, the weighted residual will contain contributions from the physical entropy produced at the shock.

Another possibility is to use the local entropy production $\Pi_i$ as the error indicator. For smooth solutions, the exact entropy evolution equation is $A\eta_t + (A\Phi)_x = 0$, which for steady flows reduces to $(A\Phi)_x = 0$. The corresponding discrete equation can be read off from (34),

$$A_{i+\frac{1}{2}}\Phi^*_{i+\frac{1}{2}} - A_{i-\frac{1}{2}}\Phi^*_{i-\frac{1}{2}} = \Pi_i \tag{35}$$

Summing over cells we get

$$A\Phi^*_{out} - A\Phi^*_{in} = \sum_i \Pi_i \tag{36}$$

The LHS is an approximation to *J*, and should be zero for the exact (analytic) solution. The RHS is thus the error in the functional, and $\Pi_i$ (production of spurious entropy) is the appropriate local error indicator. Incidentally, the left-hand side of (35) has been used as shock detector and error indicator in [14], [15], [35], etc.

Using $\Pi_i$ as the error indicator has several advantages over entropy-variable-weighted residuals. It is more clearly linked to (discrete) entropy production, and it does not require auxiliary fine meshes or higher-order reconstructions to compute the residuals. It does however depend on the numerical scheme and thus it requires some amount of hand-coding, but the required computations can be readily assembled from the subroutines used to compute the flow residuals.

In shocked cases the situation is more involved. First, entropy is not conserved but is created at the shock. The exact entropy evolution equation is now $A\eta_t + (A\Phi)_x \leq 0$ in the sense of distributions, the negative sign corresponding to production of entropy at shocks. It seems that, a priori, any indicator targetting spurious entropy creation must take that into account to avoid unnecesary refinement at shocks. The (steady) discrete equation is still (35), where now $\Pi_i$ contains both physical and spurious contributions, the former being limited to the shock region. Hence, in order to use $\Pi_i$ as the indicator for spurious entropy production, we need to either exclude the shock region from the target area or to subtract the physical entropy jump. The first possibility is relatively simple to implement in the quasi-1D case and higher dimensions, using $\Pi_i$ itself to detect the shock cells (see Fig. 5).

The second possibility is mostly restricted to the quasi-1D case, as implementation in 2D or 3D cases would be very complex. Starting from (36) we subtract from both terms the shock entropy jump

$$A\Phi^*_{out} - A\Phi^*_{in} - [A\Phi]_{x_s} = A\Phi^*_{out} - A\Phi^*_{in} - (A\Phi_{out} - A\Phi_{in}) =$$
$$\sum_i \Pi_i - (A\Phi_{out} - A\Phi_{in})$$

where $\Phi_{out}$ (resp. $\Phi_{in}$) is the analytic entropy flux function evaluated at the outlet (resp. inlet). In order to obtain a local error indicator that only targets spurious entropy production we need to subtract the physical entropy jump from $\Pi_i$ (that is, we have to distribute $(A\Phi_{out} - A\Phi_{in})$ among the nodes within the captured shock layer. To identify the nodes in the shock layer we can proceed as follows. We compute the peak value of $\Pi_i$ (which is always attained within the shock layer, see Fig. 5) and define $\sigma_i = \frac{\Pi_i}{\max \Pi_i}$. A node belongs to the shock layer if $\sigma_i > \kappa$, where $\kappa \leq 1$ is a user-defined threshold (this is somewhat arbitrary, much as shock detectors in JST dissipation [36], for example). Finally, since $\Pi_i$ should be zero away from shocks, we compute the local error indicator as

$$\varepsilon_i = |\Pi_i|$$

for non-shock cells and (37)

$$\varepsilon_i = \left| \Pi_i - \frac{\sigma_i}{\sum_{\sigma_i > \kappa} \sigma_i} (A\Phi_{out} - A\Phi_{in}) \right|$$

for shock cells.

### 3.3 Numerical tests

We consider transonic inviscid flow on a converging-diverging duct with cross section

$$A(x) = \begin{cases} 1 + \sin^2(\pi x), & |x| < 0.5 \\ 2, & 0.5 \leq |x| \leq 1 \end{cases}$$

and flow conditions such that a shock forms at $x_s \approx 0.152$ as depicted in Fig. 2, which shows the analytic solution obtained with the Area-Mach number relation [37]. The analytic entropy variables can be obtained from the flow solution using (12), while the Oswatitsch adjoint variables have been obtained in [26] using the Green's function approach of [38]. A comparison between entropy and Oswatitsch adjoint variables for shocked and shock-free cases is shown in Fig. 3. In the former case, both sets of variables show quite different trends: while entropy variables are continuous at the sonic throat and discontinuous at the shock, Oswatitsch adjoint variables are continuous at the shock with zero gradient but show a finite jump at the sonic throat. In the latter case, the Oswatitsch adjoint variables are identical to the entropy variables.

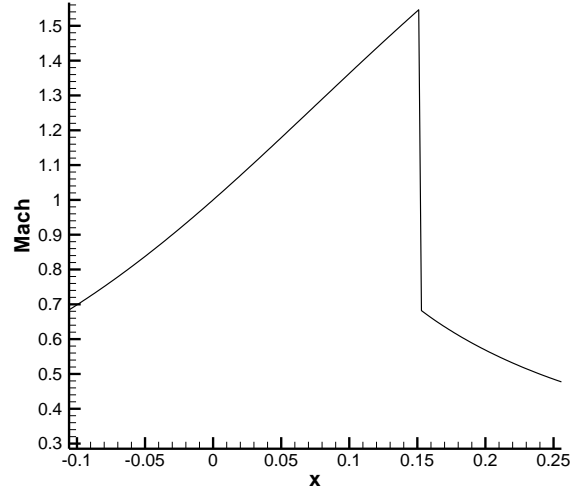

**Fig. 2 (Analytic) Mach number distribution for the transonic quasi-1D test case**

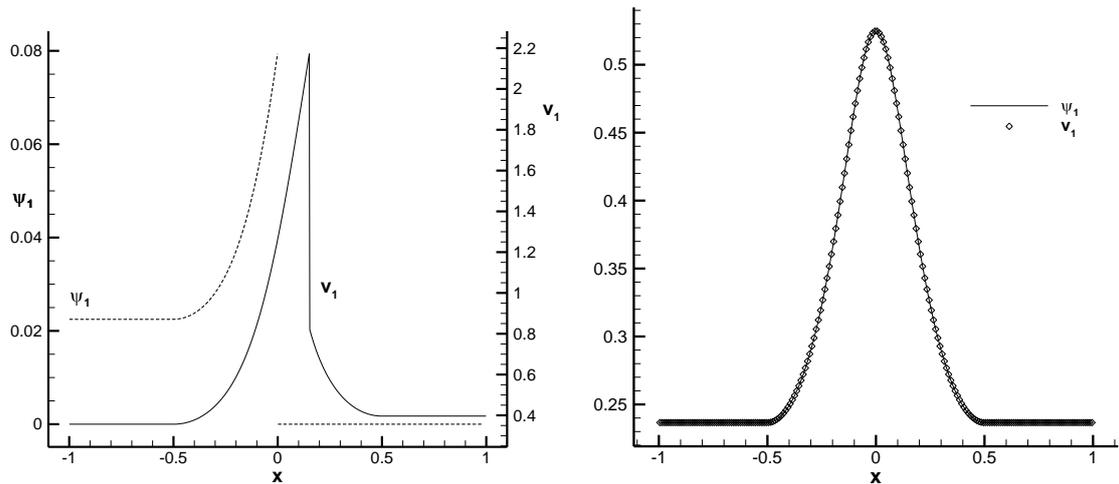

**Fig. 3 Entropy variables and Oswatitsch adjoint for shocked and shock-free ( $M_{in} = 0.2$ ) flow conditions.**

Numerical computations are carried out with a cell-centered, finite-volume discretization and three different flux functions:
- A central scheme with Jameson-Schmidt-Turkel (JST)-type artificial dissipation [36] described in [39].
- Roe's upwind scheme [40] with $1^{st}$ and $2^{nd}$ order reconstruction.
- Chandrasekhar's central Kinetic-Energy-Preserving and entropy-conservative scheme [41] (with JST-type artificial dissipation).

Fig. 4 shows the entropy and entropy variables computed with the above flux functions, while Fig. 5 shows the local value of the entropy production term $\Pi_j$. Notice that, in all cases, $\Pi_j$ is only appreciably different from zero in a layer of 3-4 cells around the captured shock, and that the peak value is attained at an intermediate state within the shock layer. Likewise, although not shown, the height of the peak remains approximately the same regardless of the mesh spacing, while its width shrinks as the mesh is refined.

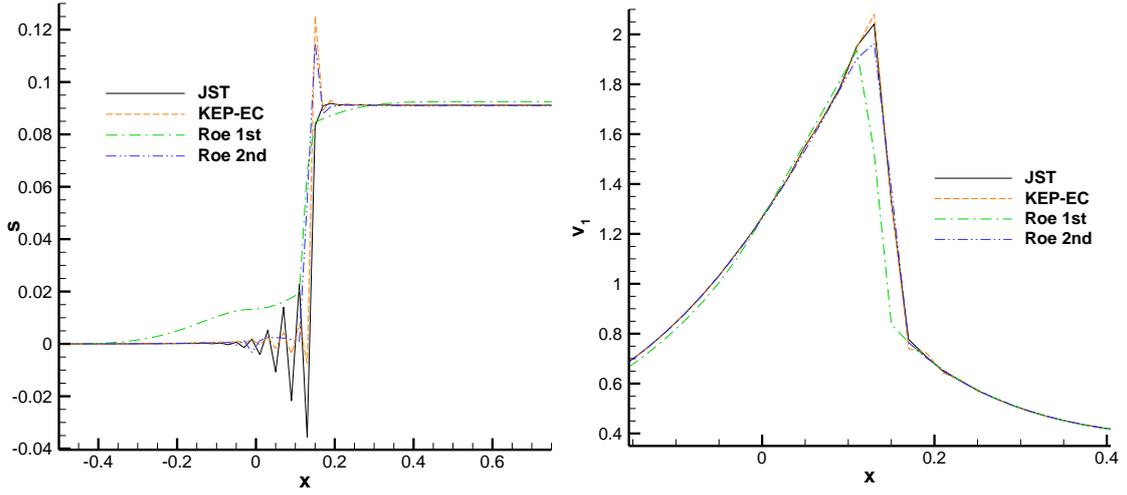

**Fig. 4.** Plot of $s/R = \log(p/\rho^\gamma)/(\gamma-1)$ and $v_1 = \dfrac{\rho u}{p}$

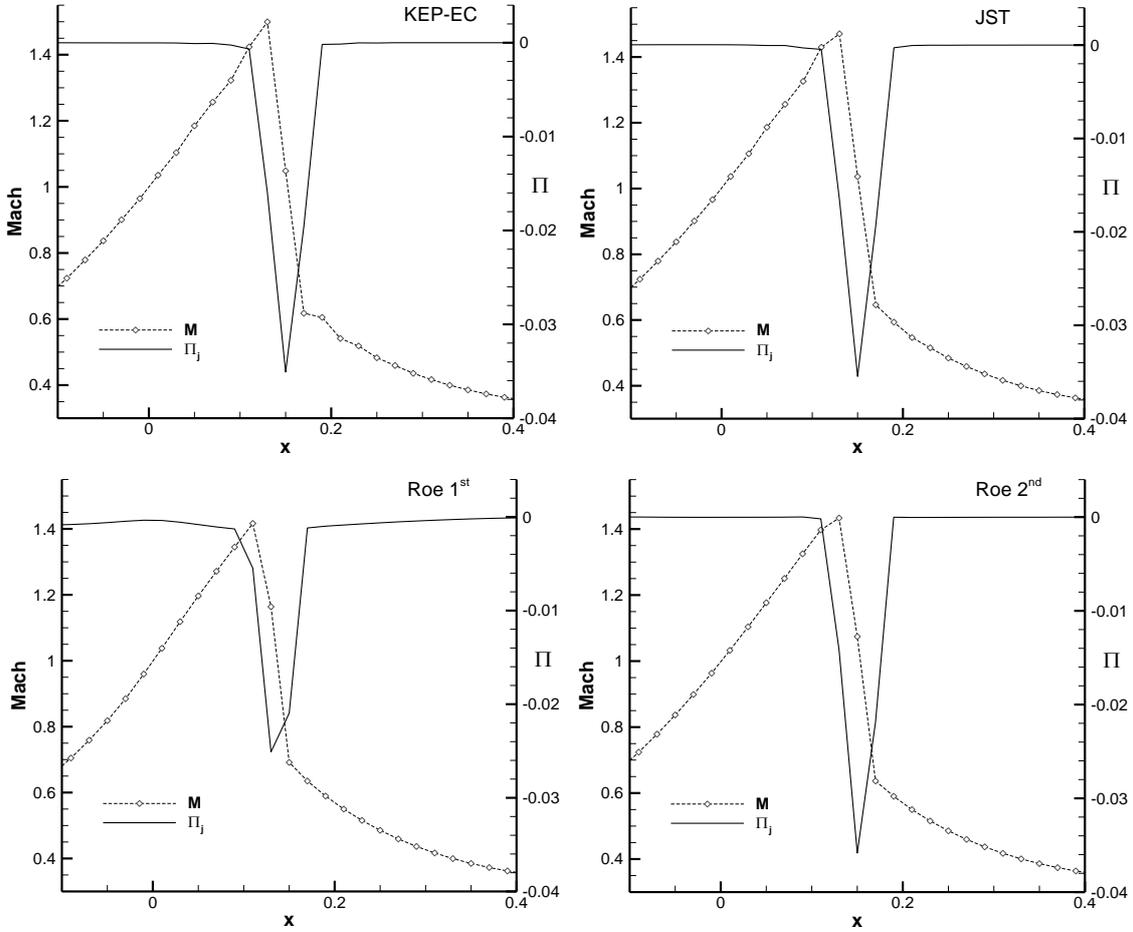

**Fig. 5.** Plot of $\Pi_j = \left(\Pi_{j+1/2} + \Pi_{j-1/2}\right)/2$ **for the steady-state shocked quasi-1D solution computed with different flux functions. Mach number distribution is shown for reference.**

Next, we perform an entropy-production based mesh adaptation (actually, refinement) exercise with the JST scheme. In a first analysis, the adaptation is carried out by splitting in half the cells according to the value of the entropy production $\Pi_i$ with modifications following the strategies outlined above: (1) completely exclude cells in the shock layer, (2) subtract the physical entropy jump from $\Pi_i$ at shock points, and (3), leave $\Pi_i$ unchanged. Shock cells are those for which $\sigma_i = \frac{\Pi_i}{\max \Pi_i} > 0.1$. The criterion to decide whether a cell is flagged for division is based on a threshold value that is set implicitly by requiring that the percentage of new cells is fixed at 40%.

In order to assess the quality of the resulting meshes, we evaluate the "lift" integral $J_L = \int_{-1}^{1} p dx$. A convergence plot comparing the three strategies is shown in Fig. 6 (a), where comparison with isotropic refinement (whereby each cell is split in two at each stage) is also provided. It can be seen that strategies (2) and (3) produce nearly identical convergence plots for $J_L$ and likewise quite similar meshes (Fig. 7). Excluding the shock layer from refinement not only produces odd-looking adapted meshes, but also yields wrong values for $J_L$.

The adaptation strategy based on the local entropy production is further assessed in Fig. 6 (b) by comparison with three additional adjoint-based adaptation procedures: the entropy adjoint (i.e., the entropy variables), the "lift" adjoint (i.e., the adjoint relative to $J_L$) and the Oswatitsch adjoint (i.e., the adjoint relative to the inlet/exit entropy balance $[A\Phi]_{-1}^{+1}$). The lift and Oswatitsch adjoint solutions have been obtained with a continuous adjoint solver described and tested in [39, 26] that uses a finite-volume central scheme with JST artificial dissipation. In these three cases, the local adaptation indicator is built from the corresponding adjoint solution $\psi$ using the dissipation part of the JST residual, $R^{(diss)}$, following Dwight's formulation [33], as

$$\eta_i = \left| \psi^T \left( k_2^i \frac{\partial R^{(diss)}}{\partial k_2^i} + k_4^i \frac{\partial R^{(diss)}}{\partial k_4^i} \right) \right| \qquad (38)$$

where $k_2^i$ and $k_4^i$ are local values of the dissipation coefficients. This indicator focuses on the part of the discretization error that can be traced to the artificial dissipation (thus missing other potentially significant contributions to the total error), but it has the advantage of only requiring a single mesh in contrast to other approaches such as [29]. All other mesh adaptation settings are exactly as described above for the entropy production-based adaptation approach. Fig. 6 (b) also shows the adjoint-based corrected "lift" values obtained with the lift adjoint solution as

$$J_L^{corr} = J_L + \sum_{cells} \psi^T \left( k_2 \frac{\partial R^{(diss)}}{\partial k_2} + k_4 \frac{\partial R^{(diss)}}{\partial k_4} \right) = J_L + \sum_{cells} \psi^T R^{(diss)}$$

The error is fixed with respect to a reference value obtained with Richardson extrapolation from the isotropic refinement results.

We see that all adaptation indicators perform comparably, with both entropy-based strategies, which follow similar trends, outperforming the lift-based adaptation in the long term. The Oswatitsch adjoint, on the other hand, follows more closely the lift adjoint scheme. This behaviour can be confirmed with the adapted meshes, where entropy adjoint and entropy production schemes tend to refine more intensely the shock region, while the lift adjoint and the Oswatitsch adjoint concentrate the refinement in the throat region. It is particularly striking that, in general, all estimators tend to saturate rather quickly and, in fact, most of them are most accurate on the initial grids. As explained in [33], this is a characteristic of dissipation-based adaptation, which is very effective in removing the part of the error due to dissipation, with the remaining error not due to dissipation eventually dominating as the adaptation progresses.

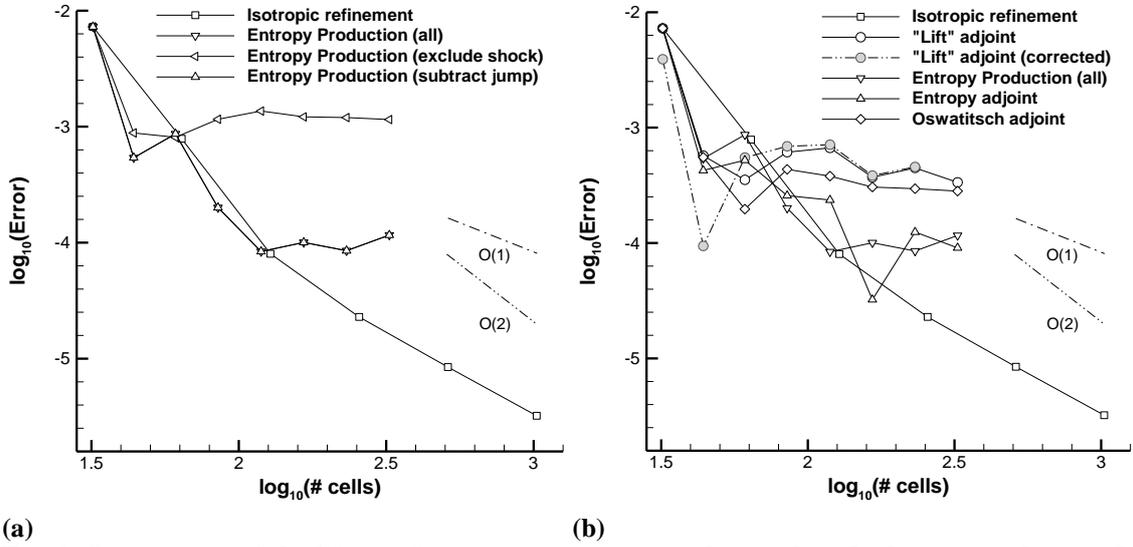

**Fig. 6.** Convergence of the lift cost function. (a) entropy production-based adaptation schemes. (b) Comparison of different adaptation schemes.

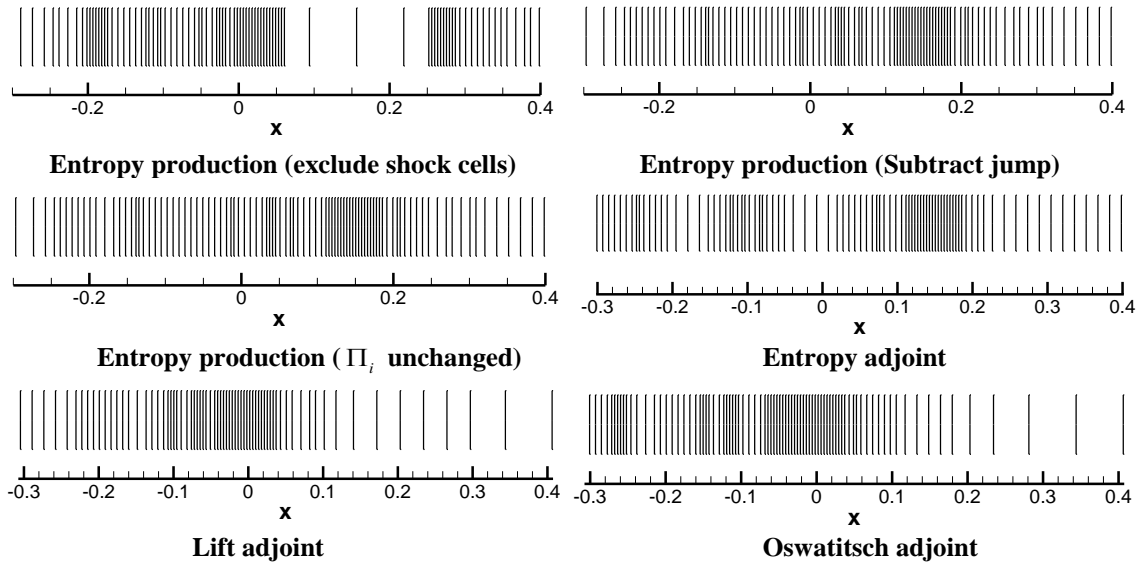

**Fig. 7.** Adapted meshes after 4 adaptation iterations with different adaptation strategies.

## 4  2D Euler equations

The 2D Euler equations in conservation form are

$$R(U) = \nabla \cdot \vec{F} = 0 \tag{39}$$

where $U$ is the vector of conservation variables and $\vec{F} = (F_x, F_y)$ is the vector flux function

$$U = \begin{pmatrix} \rho \\ \rho u_x \\ \rho u_y \\ \rho E \end{pmatrix}, \qquad F_x = \begin{pmatrix} \rho u_x \\ \rho u_x^2 + p \\ \rho u_x u_y \\ \rho u_x H \end{pmatrix}, \qquad F_y = \begin{pmatrix} \rho u_y \\ \rho u_x u_y \\ \rho u_y^2 + p \\ \rho u_y H \end{pmatrix} \qquad (40)$$

**4.1 Continuous analysis**

In 2D, the entropy is again conserved away from shocks, where it jumps. The entropy function can be again chosen simply as

$$\eta = -\rho s / R, \qquad (41)$$

This choice yields entropy variables

$$\mathbf{v}^T = \left( \frac{\gamma}{\gamma - 1} - \frac{s}{R} - \frac{\rho \vec{u}^2}{2p}, \frac{\rho \vec{u}}{p}, -\frac{\rho}{p} \right) \qquad (42)$$

and entropy flux

$$\vec{\Phi} = \vec{u}\eta = -\rho \vec{u} s / R \qquad (43)$$

which satisfies the compatibility condition

$$\mathbf{v}^T \frac{\partial \vec{F}}{\partial U} = \frac{\partial \vec{\Phi}}{\partial U} \qquad (44)$$

Multiplying the Euler equations (39) by $\mathbf{v}^T$, and using (44), the entropy conservation law is recovered

$$\mathbf{v}^T R(U) = \nabla \cdot \vec{\Phi} = 0 \qquad (45)$$

Likewise, writing the flow equations in terms of the entropy variables and using their symmetry properties, it follows that the entropy variables satisfy an adjoint-like equation

$$\vec{F}_U^T \cdot \nabla \mathbf{v} = 0 \qquad (46)$$

Integrating the right-hand side of (45) on the fluid domain $\Omega$ yields the following relation

$$0 = \int_\Omega \mathbf{v}^T R(U) d\Omega = \int_\Omega \nabla \cdot \left( -\frac{\rho \vec{u} s}{R} \right) d\Omega = -\int_{S_\infty} \frac{\rho \vec{u} s}{R} \cdot \hat{n} dS \qquad (47)$$

in terms of the integrated entropy flux across the far-field $S_\infty$, where the flow configuration sketched in Fig. 1 has been assumed.

Eq. (47) can also be generalized to flows with shocks. If the solution $U$ contains a shock along a curve or surface (in 3D) $\Sigma$, the Rankine-Hugoniot jump condition $[\vec{F}] \cdot \hat{n}_\Sigma = 0$ connects the smooth solutions $R(U) = 0$ on either side. Here $[\vec{F}] = \vec{F}|_{downstream} - \vec{F}|_{upstream}$ denotes the jump in $\vec{F}$ across the shock and $\hat{n}_\Sigma$ is the unit normal vector to $\Sigma$ (oriented such that it points upstream of the shock). Likewise, eq. (45) also remains valid on either side of the shock, in such a way that integrating the right-hand side of (45) yields the following relation[1]

$$0 = \int_{\Omega\setminus\Sigma} v^T R(U) d\Omega = \int_{\Omega\setminus\Sigma} \nabla \cdot \left( -\frac{\rho \vec{u} s}{R} \right) d\Omega = -\int_{S_\infty} \frac{\rho \vec{u} s}{R} \cdot \hat{n} dS - \int_{\Sigma} \frac{[\rho \vec{u} s]_\Sigma}{R} \cdot \hat{n}_\Sigma d\Sigma \qquad (48)$$

As in the quasi-1D case, and taking (46) into account, it is possible to show that the entropy variables are adjoint variables relative to the output functions (47) (for flows without shocks, see also [42]) and (48) (for shocked flows). In the former case, the target functional is essentially Oswatitsch's formula for aerodynamic drag as pointed out in [17].

When the flow contains shocks, the entropy variables are different from the corresponding Oswatitsch adjoint solution (see below), but they can still be used to target spurious entropy production throughout the domain, the reduction of which through mesh adaptation will likely improve the quality of the solution as well as the prediction of global functionals such as drag or lift. In the 2D case, the entropy is constant along each streamline (with entropy possibly varying between streamlines depending upon the incident state) except at shocks, where it jumps. The entropy jump at the shock is generically different for each streamline. Hence, subtracting the physical entropy jump from the entropy production sensor as we did in the quasi-1D case is, for all practical purposes, impossible. Alternatives that combine different sensors, such as the masked sensor in [19], will be explored below.

As in the quasi-1D case, in the shocked case it is possible to build the Oswatitsch adjoint –the adjoint corresponding to the far-field entropy flux

$$J = \int_{\partial\Omega} \vec{\Phi} \cdot \hat{n} dS = -\frac{1}{R} \int_{\partial\Omega} \rho s \vec{u} \cdot \hat{n} dS = -\frac{1}{R} \int_{S_\infty} \rho s \vec{u} \cdot \hat{n} dS$$

Details can be found in [42]. The derivation follows the approach outlined in [43] (see also [26]) for a functional consisting on a function of the pressure integrated along the wall and can be easily adapted to the present case. The Oswatitsch adjoint $\psi$ obeys the adjoint equation

$$\nabla \psi^T \cdot \vec{F}_U = 0 \qquad \text{in } \Omega\setminus\Sigma \qquad (49)$$

with the following wall and far-field boundary conditions

$$\vec{\varphi} \cdot \hat{n} = 0 \qquad \text{on } S_w \setminus x_b$$
$$(v - \psi)^T (\vec{F}_U \cdot \hat{n}) \delta U = 0 \qquad \text{on } S_\infty \qquad (50)$$

($x_b = \Sigma \cap S_w$ is the shock foot). Furthermore, $\psi$ must be continuous across the shock

$$[\psi]_\Sigma = 0 \qquad (51)$$

where it must obey an internal differential equation

---

[1] Transonic rotational flows contain an additional singularity in the form of a slip line emanating from the sharp trailing edge. However, this is a contact discontinuity (no mass flow), so it does not contribute to (48)

$$\partial_{tg}\psi^T[\vec{F}\cdot\vec{t}_\Sigma]_\Sigma = 0 \tag{52}$$

along the shock, and

$$\psi^T(x_b)[\vec{F}\cdot\vec{t}_\Sigma]_{x_b} = 0 \tag{53}$$

at the shock foot $x_b$ ($[\ ]_{x_b}$ is the jump across the shock at the shock foot). As explained in [26], the behavior of the adjoint normal derivatives at the shock can be investigated using the adjoint equations, which at the shock yield $[\vec{F}_U^T\cdot(\hat{t}_\Sigma\partial_{tg}\psi+\hat{n}_\Sigma\partial_n\psi)]_\Sigma = 0$, and the shock equations (52) and (53). In this way, it can be shown that for normal shocks, for example, adjoint normal derivatives are mostly vanishing (and continuous) across the shock. At any rate, these equations are usually not taken into account for the discretization of the adjoint equations. It is assumed that the discretized adjoint system picks the correct solution satisfying (52) and (53) provided that there is enough dissipation across the shock [44]. However, taking the shock relations into account does have an impact in shape optimization [43].

### 4.2 Discrete analysis

Let us now try to find a suitable discrete representation of the entropy adjoint construction for mesh adaptation purposes. To fix ideas, let us assume that the equations (39) are numerically approximated on unstructured grids using a node-centered, edge-based central finite-volume discretization on a dual grid (see Barth [45] for further details) with artificial dissipation. The resulting semidiscrete scheme can be written as

$$|\Omega_i|\frac{dU_{h,i}}{dt}+R_{h,i}(U_h) = 0 \tag{54}$$

where, for interior nodes,

$$R_{h,i} = \sum_{j:i}\vec{n}_{ij}\cdot\vec{f}_{ij}^* + d_{ij} \tag{55}$$

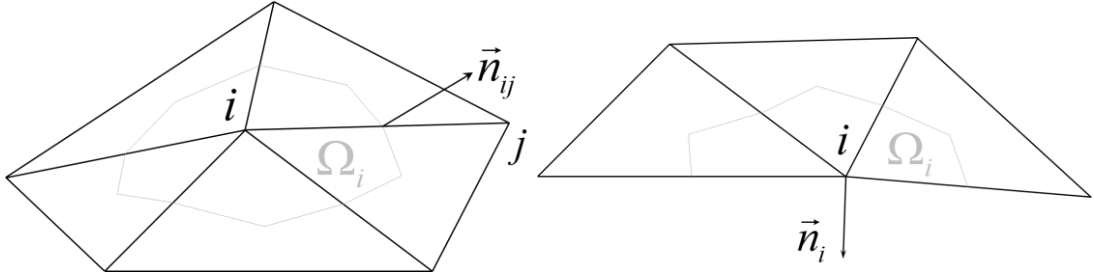

**Fig. 8. Dual grid cell associated with interior node $i$ and integrated normal vector $\vec{n}_{ij}$ associated with edge $ij$ (left). Dual grid cell and integrated normal vector associated with boundary node $i$ (right).**

Here $\Omega_i$ is the dual grid cell (with area/volume $|\Omega_i|$) associated to a node $i$, $j:i$ denotes the set of nodes $j$ that are connected to node $i$ by an edge, $\vec{n}_{ij}$ is the integrated normal vector of the dual grid face associated with edge $ij$ (see Fig. 8) obeying $\sum_{j:i}\vec{n}_{ij} = 0$, while the numerical fluxes associated with edge $ij$ are taken as $\vec{f}_{ij}^* = (\vec{F}_i+\vec{F}_j)/2$, where $\vec{F}_i = \vec{F}(U(\vec{x}_i))$ is the flux vector (40) at cell $i$, resulting in a central scheme. Finally, the

artificial dissipation flux $d_{ij}$ is typically a blend of second and fourth-order differences [36]

$$d_{ij} = \varepsilon_{ij}^{(2)}(U_i - U_j) + \varepsilon_{ij}^{(4)}(\nabla^2 U_i - \nabla^2 U_j) \tag{56}$$

with $\nabla^2 U_i = \sum_{k:i}(U_k - U_i)$ and shock switches $\varepsilon_{ij}^{(2)}$ and $\varepsilon_{ij}^{(4)}$. On boundary nodes, (55) is replaced with

$$R_{h,i} = \sum_{j:i}(\vec{n}_{ij} \cdot \vec{f}_{ij}^* + d_{ij}) + \vec{n}_i \cdot \vec{f}_i^{bc} \tag{57}$$

where $\vec{f}_i^{bc}$ are the boundary fluxes and $\vec{n}_i$ is the normal vector to the boundary face obeying $\vec{n}_i = -\sum_{j:i}\vec{n}_{ij}$. Multiplying (54) by the entropy variables and rearranging yields the discrete entropy evolution equation. For interior nodes one has

$$|\Omega|_i \mathrm{v}_{h,i}^T \frac{dU_{h,i}}{dt} + \mathrm{v}_{h,i}^T R_{h,i} = |\Omega|_i \frac{d\eta_{h,i}}{dt} + \sum_{j:i}\vec{n}_{ij} \cdot \vec{\Phi}_{ij}^* - \sum_{j:i}\Pi_{ij} = 0 \tag{58}$$

where $\vec{\Phi}_{ij}^* = \overline{\mathrm{v}}_{ij}^T(\vec{f}_{ij}^* + \frac{1}{\vec{n}_{ij}\cdot\vec{n}_{ij}}\vec{n}_{ij}d_{ij}) - \overline{\vec{\Theta}}_{ij}$ is the entropy flux and

$$\Pi_{ij} = \frac{1}{2}\left(\Delta \mathrm{v}_{ij}^T(\vec{n}_{ij}\cdot\vec{f}_{ij}^* + d_{ij}) - \vec{n}_{ij}\cdot\Delta\vec{\Theta}_{ij}\right) \tag{59}$$

is the entropy production. In the above equations, $\vec{\Theta}_i = \rho_i \vec{u}_i$ is the entropy potential, and the notation $\overline{(\cdot)}_{ij} = \frac{1}{2}((\cdot)_j + (\cdot)_i)$ and $\Delta(\cdot)_{ij} = (\cdot)_j - (\cdot)_i$ has been used. On boundary nodes, the semidiscrete entropy evolution equation is

$$|\Omega|_i \frac{d\eta_{h,i}}{dt} + \sum_{j:i}\vec{n}_{ij}\cdot\vec{\Phi}_{ij}^* - \sum_{j:i}\Pi_{ij} + \vec{n}_i\cdot(\mathrm{v}_{h,i}^T\vec{f}_i^{bc} - \vec{\Theta}_i) = 0 \tag{60}$$

Notice that if the boundary fluxes are evaluated as $\vec{f}_i^{bc} = \vec{F}(U(\vec{x}_i))$, then the boundary entropy flux $\vec{n}_i\cdot(\mathrm{v}_i^T\vec{f}_i^{bc} - \vec{\Theta}_i)$ is exactly $\vec{n}_i\cdot\vec{\Phi}(U(\vec{x}_i))$. For inviscid flows, $\vec{n}_i\cdot\vec{u}_i = 0$ and $\vec{f}_i^{bc} = (\vec{0}, \mathrm{I}p_i, 0)^T$ at solid boundaries and, thus, $\mathrm{v}_i^T\vec{f}_i^{bc}\cdot\vec{n}_i = (\rho\vec{u})_i\cdot\vec{n}_i = 0$ and $\vec{n}_i\cdot\vec{\Theta}_i = (\rho\vec{u})_i\cdot\vec{n}_i = 0$, so there is no entropy flux across walls. The boundary entropy flux is only non-vanishing for the far-field boundary, where $\vec{f}_i^{bc}$ is usually given in terms of characteristic variables.

Summing (58) and (60) over the (dual) mesh cells yields the evolution equation for the integrated entropy

$$\sum_{i\in I\cup\partial I}|\Omega|_i \frac{d\eta_{h,i}}{dt} + \sum_{i\in I\cup\partial I}\sum_{j:i}\vec{n}_{ij}\cdot\vec{\Phi}_{ij}^* + \sum_{i\in\partial I}\vec{n}_i\cdot(\mathrm{v}_i^T\vec{f}_i^{bc} - \vec{\Theta}_i) = \sum_{i\in I\cup\partial I}\sum_{j:i}\Pi_{ij} \tag{61}$$

where $I$ denotes the set of all interior mesh nodes and $\partial I$ the set of all boundary nodes. The second term on the left-hand side is zero since it is the mesh sum of a conservative flux ($\vec{\Phi}_{ij}^*$ is symmetric in $ij$, and thus $\vec{n}_{ij}\cdot\vec{\Phi}_{ij}^*$ is antisymmetric in $ij$). Hence, for converged steady flows the above equation yields the discrete entropy conservation equation

$$\sum_{i\in\partial I}\vec{n}_i\cdot(\mathrm{v}_i^T\vec{f}_i^{bc}-\vec{\Theta}_i) = \sum_{i\in I\cup\partial I}\sum_{j:i}\Pi_{ij} \qquad (62)$$

The LHS is the discrete form of the boundary entropy flux $\int_{\partial\Omega}\hat{n}\cdot\vec{\Phi}ds$, while the RHS is the integrated entropy production throughout the mesh. As in the quasi-1D case, it is clear from Eq. (62) that for smooth flows $\sum_{i\in I\cup\partial I}\sum_{j:i}\Pi_{ij}$ measures the error in entropy conservation and $\sum_{j:i}\Pi_{ij}$ can thus serve as local mesh adaptation indicator for cell $i$. When shocks are present, $\sum_{j:i}\Pi_{ij}$ also contains the physical production of entropy at the shocks.

### 4.3 Numerical tests

The first example is inviscid, subcritical flow over a NACA 0012 airfoil with free-stream Mach number $M_\infty=0.5$ and angle of attack $\alpha=2^\circ$. This case has been addressed in [17], so we will not examine it in detail here. Instead, we will only compute the entropy and Oswatitsch adjoint variables and compare them on the airfoil surface. This is done in Fig. 9, showing a good agreement which confirms that, as in the quasi-1D case, the entropy variables are dual to the Oswatitsch functional.

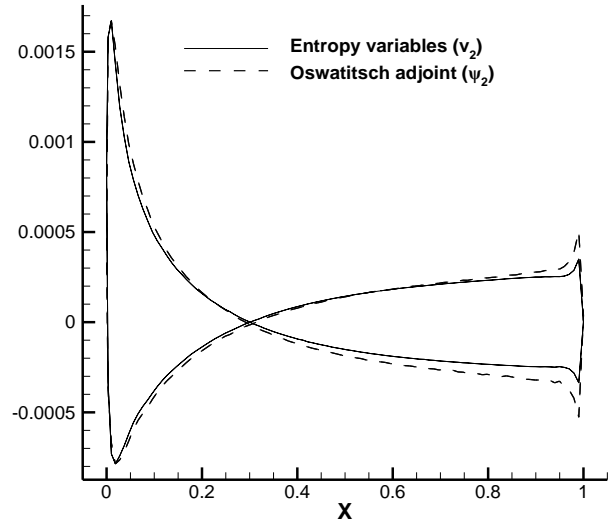

**Fig. 9. NACA0012 $M_\infty=0.5$ and $\alpha=2^\circ$. Comparison between entropy variables and Oswatitsch adjoint on the airfoil surface.**

In this case and in what follows, the flow and adjoint computations have been carried out with DLR's unstructured, finite volume solver TAU [46]. Both the flow and adjoint solvers use a cell-vertex, second-order, central discretization with JST scalar dissipation [36] of the type described in section 4.2. The drag/lift and Oswatitsch adjoint solutions are computed with the same continuous adjoint solver with only minor modifications affecting the boundary conditions. In both cases, the following adjoint boundary conditions

$$\vec{\varphi}\cdot\hat{n}=\vec{f}\cdot\hat{n} \qquad \text{on } S_w$$
$$\chi^T(\vec{F}_U\cdot\hat{n})=0 \qquad \text{along outgoing characteristics on } S_\infty$$

are weakly enforced, with $\chi = \psi$ and $\vec{f}$ equal to the force direction for drag/lift adjoints, and $\chi = \psi - \mathrm{v}$ and $\vec{f} = 0$ for the Oswatitsch adjoint.

The next example concerns transonic flow past a NACA0012 airfoil with $M_\infty = 0.8$ and $\alpha = 1.25º$. The flow has a strong shock at 65% chord on the suction side of the airfoil (Fig. 10). The initial mesh, which is shown in Fig. 10, has 2995 nodes, with the far-field located approximately 50 chord-lengths away from the airfoil.

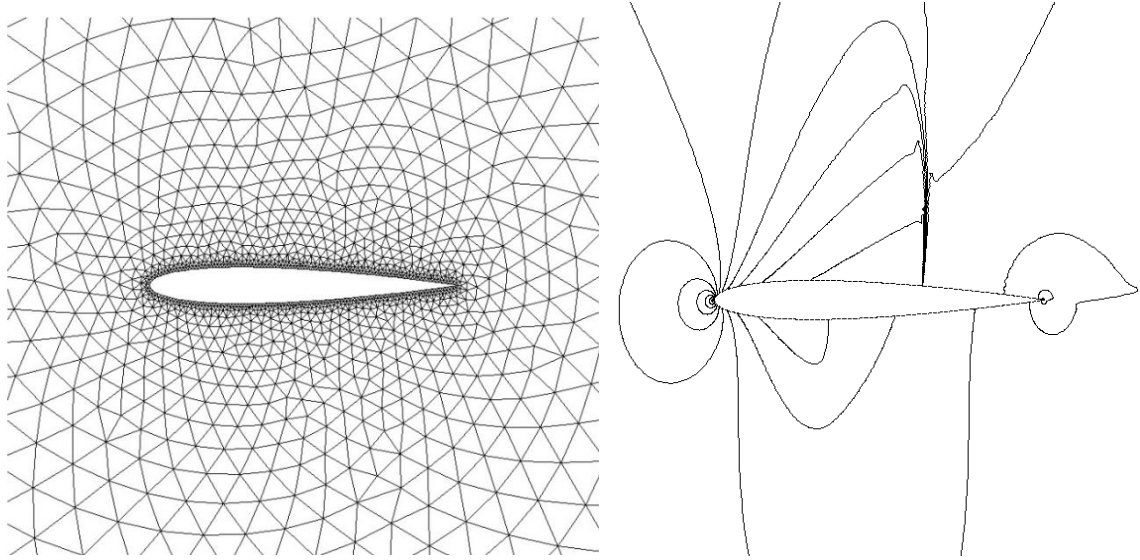

**Fig. 10 NACA0012. Initial mesh and sample numerical flow solution**

Mesh adaptation is carried out by splitting in half the mesh edges according to the value of the local adaptation indicator, in such a way that an edge is marked for refinement if the value of the indicator exceeds a threshold that is set implicitly by requiring that the percentage of new points introduced is fixed at 40%. When edges are split on the airfoil surface, the position of the new points is adjusted using cubic splines interpolation such that the new discrete surface follows the original geometry.

We shall consider different adaptation indicators: local entropy production ($\sum_{j:i} \Pi_{ij}$), as well as entropy adjoint, Oswatitsch adjoint and conventional drag and lift adjoint-based adaptation, all with Dwight's indicator (38). We also consider adaptation sensors that combine entropy production and drag-adjoint Dwight's sensor. The first one is essentially the masked sensor proposed in [19], whereby entropy production-based adaptation is applied only to a subset of cells (50% of the original cells) previously flagged for drag-adjoint based refinement. This correction pushes the entropy based adaptation towards a more engineeringly-oriented outcome as well as may help alleviate over-refinement in the shock region. Two other combined sensors follow this idea, one by considering the sum of the normalized drag-adjoint and entropy production sensors, the other by considering the point-wise product of both sensors.

Fig. 11 compares the different combined sensors with the raw entropy production sensor in terms of the convergence of the drag coefficient $c_d = \frac{2}{\rho_\infty \bar{u}_\infty^2} \oint_{S_w} (\hat{n}_x \cos\alpha + \hat{n}_y \sin\alpha) p\, dS$. Isotropic refinement (where every mesh edge is split in two) is also included for reference purposes. While combining the entropy-based sensor with the drag-adjoint-based sensor does produce some improvement, particularly at the earliest adaptation cycles, the values do not differ much from what can be obtained with the raw entropy production sensor, which has the added benefit of not requiring a separate adjoint solution.

Fig. 12 compares the convergence of drag and lift coefficients for the different adaptive strategies. Errors are computed with respect to reference values obtained by Richardson extrapolation from the isotropic refinement values. The entropy-adjoint approach yields worse results in terms of the drag coefficient than the entropy production approaches, which clearly outperform isotropic refinement and compare favorably (particularly the combined product sensor) with the specific uncorrected drag-adjoint-based result (adjoint-corrected values are also shown as filled circles). Notice that the Oswatitsch adjoint follows very closely the near-field drag adjoint-based result. In terms of lift, both entropy-based approaches go astray at the first adaptation stage (as does the corrected lift-based strategy), but quickly recover, with the entropy production approach yielding better results at intermediate stages than the entropy adjoint, both converging to the same value at later stages. The Oswatitsch adjoint, on the other hand, yields surprisingly good results, especially at intermediate stages, where it outperforms the corrected lift-based adjoint results.

As shown in Fig. 13, entropy-adapted meshes are also quite similar, though significantly different from drag-based and Oswatitsch adjoint adapted meshes, which show similar trends including intense refinement along the incoming stagnation streamline and the contour of the supersonic bubble on the suction side of the airfoil, which are typical features of adjoint-adapted meshes. The entropy-based approaches, on the other hand, clearly target the wake and shock regions, where the local entropy sensors dominate, as can be seen in Fig. 14[2]. This over-refinement does certainly have an impact on drag coefficient error convergence, especially at the earliest stages. The combined approach yields better results, but it requires an additional adjoint solution at each cycle, which can incur in significant costs especially for complex 3D cases. A possible way to combine both approaches is through an alternate strategy, whereby each adaptation cycle is alternatively performed with either approach (one entropy production-based, the next drag-adjoint-based). This approach halves the cost in adjoint numerical computations, but it does not result in significant improvements in this case.

---

[2] Notice also from Fig. 14 that, as in the quasi-1D case, entropy-based local indicators serve quite effectively as well as shock detectors.

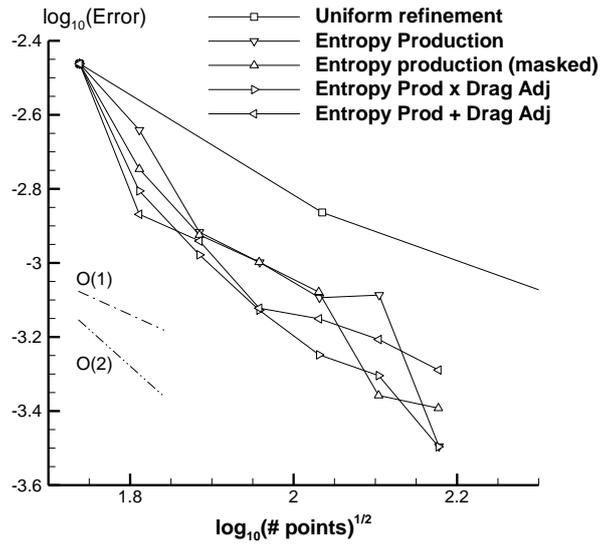

**Fig. 11.** NACA0012 $M_\infty = 0.8$ and $\alpha = 1.25°$. Convergence of drag coefficient for the entropy-production-based adaptation schemes.

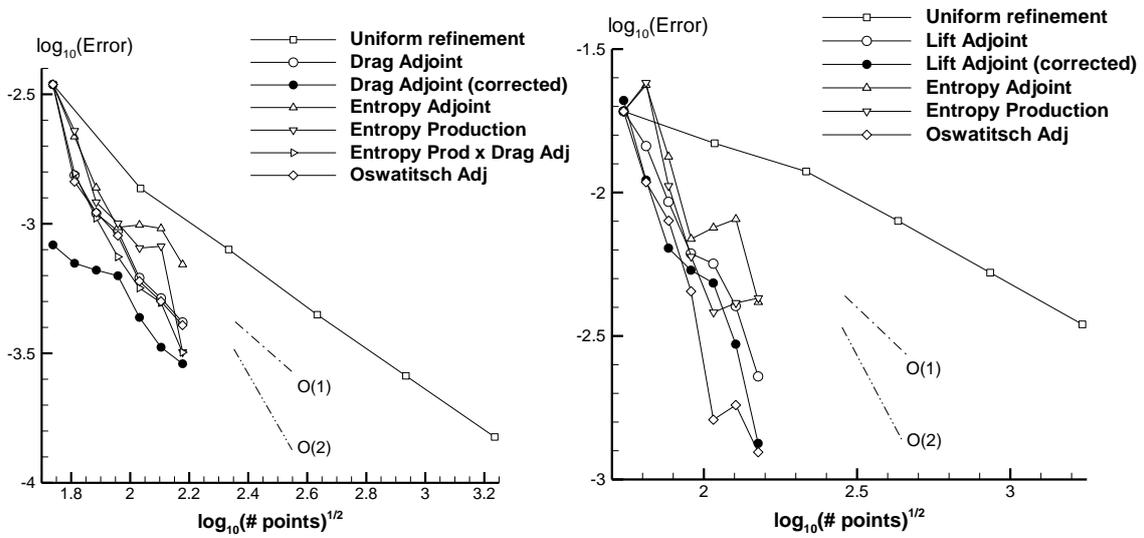

**Fig. 12.** NACA0012 $M_\infty = 0.8$ and $\alpha = 1.25°$. Convergence of drag (left) and lift (right) coefficients for different adaptation schemes.

**Entropy adjoint**

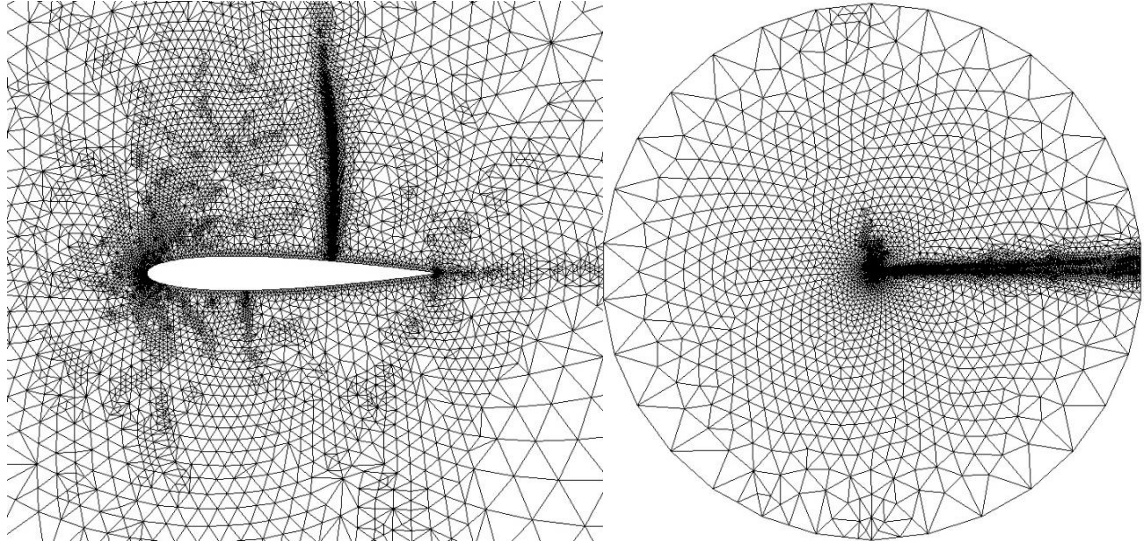

**Entropy Production**

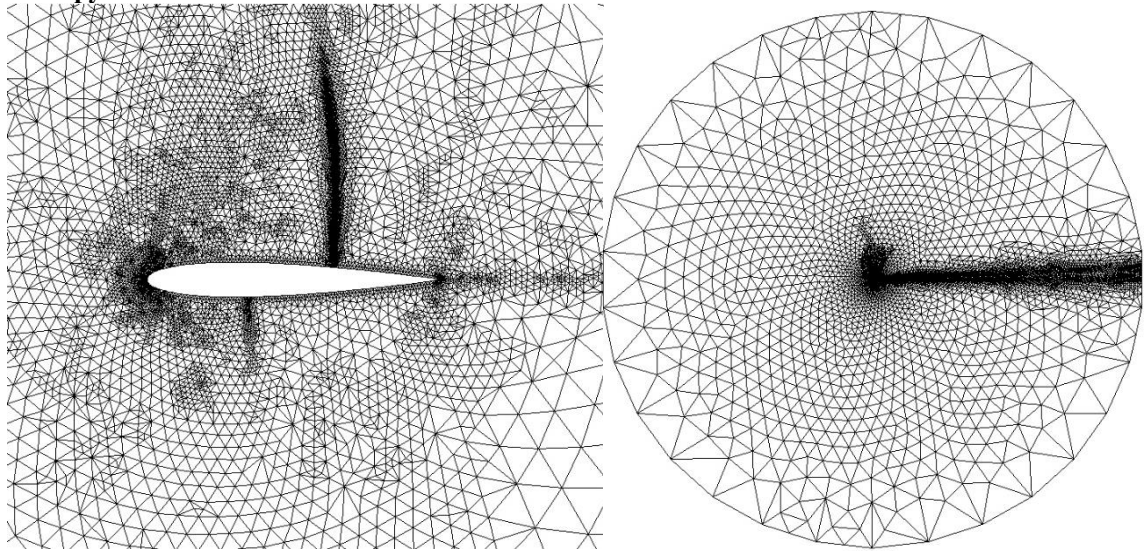

**Entropy Production × Drag adjoint**

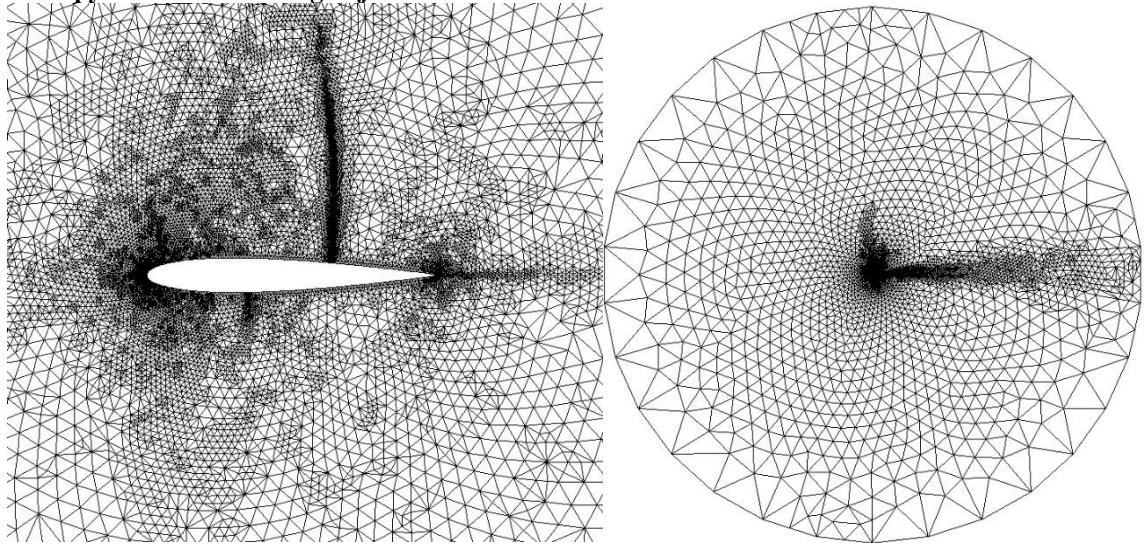

**Oswatitsch adjoint**

**Drag adjoint**

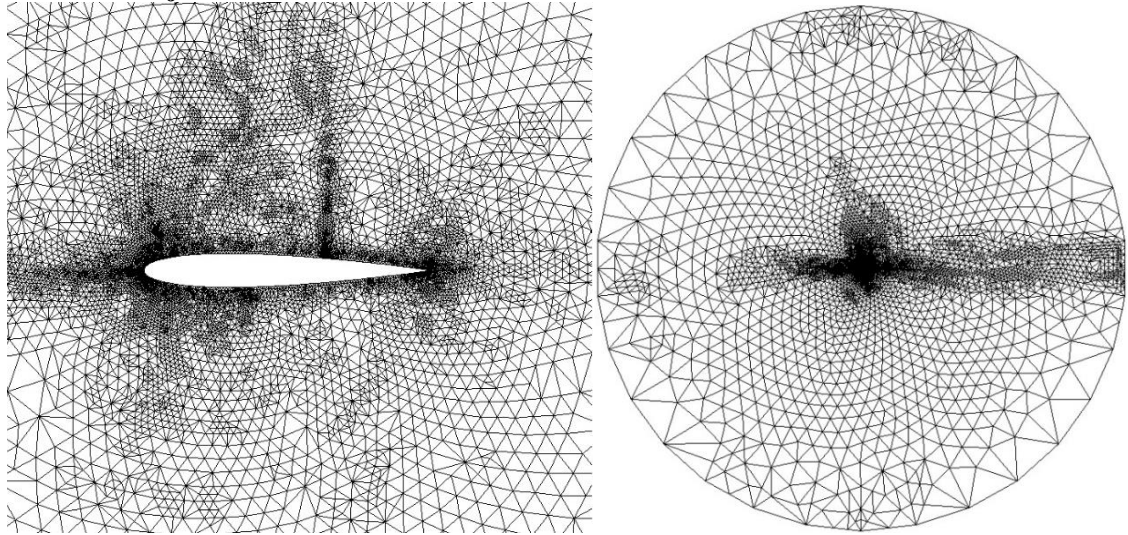

**Fig. 13.** NACA0012 $M_\infty = 0.8$ and $\alpha = 1.25°$. Final adapted meshes after 6 adaptation cycles.

**Entropy adjoint**                          **Entropy Production**

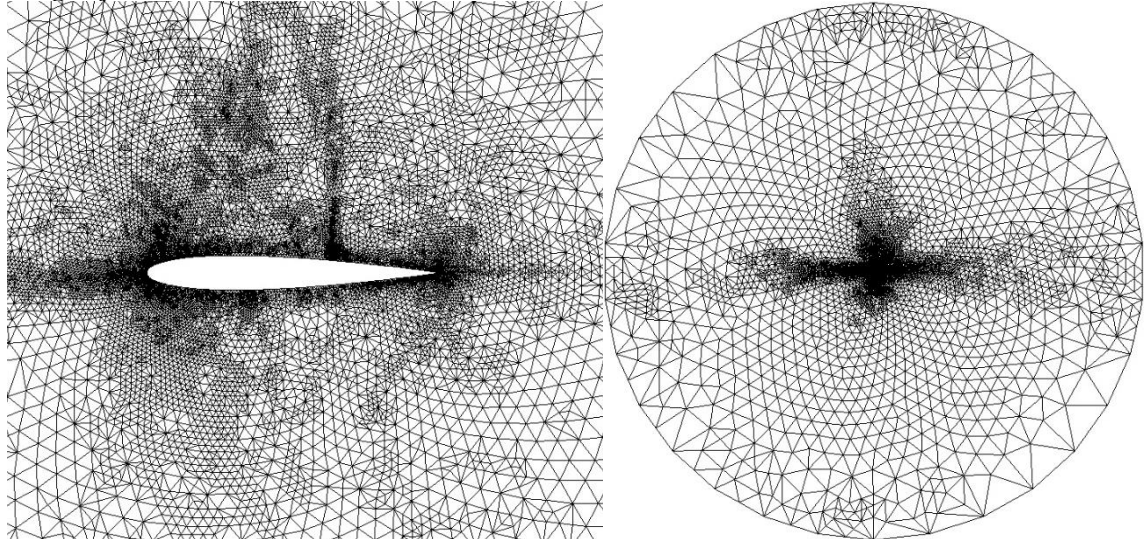

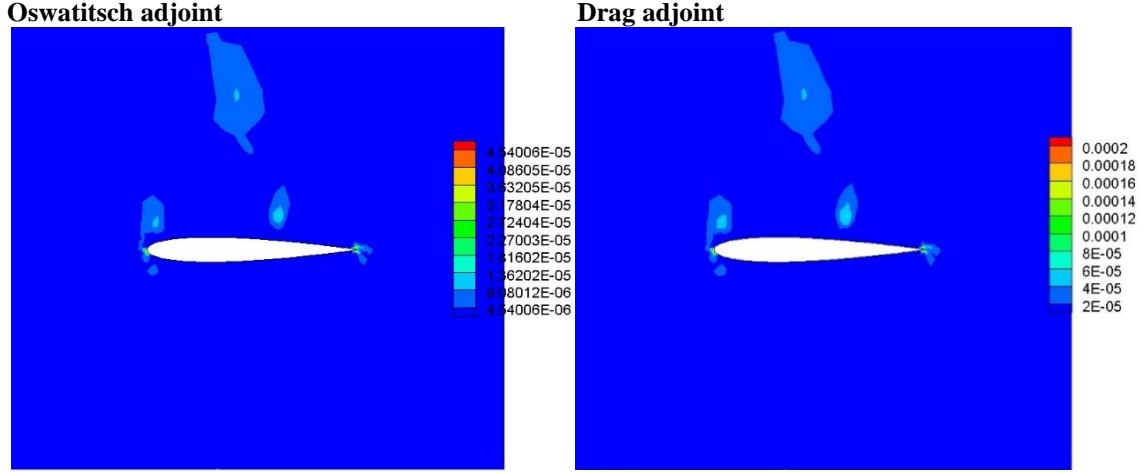

**Fig. 14. NACA0012** $M_\infty = 0.8$ **and** $\alpha = 1.25°$. **Plot of local adaptation sensors on the baseline mesh.**

## 5    3D Euler equations

The 2D analysis of the entropy and Oswatitsch adjoint approaches carries over with only the obvious modifications to 3D. Without shocks, the entropy and Oswatitsch adjoint variables are identical, while they differ in the shocked case. The Oswatitsch adjoint variables are continuous across the shock, where they obey the shock equations

$$\nabla_{tg}\psi^T \cdot [\vec{F}]_\Sigma = 0 \qquad (63)$$

on the shock surface, which is taken to be a single sheet attached to the wing surface along a curve $\sigma_S$, and

$$\psi^T[\vec{F}\cdot\hat{n}_{\sigma_S}]_{\sigma_S} = 0 \qquad (64)$$

along the shock foot $\sigma_S$ ($[\ ]_{\sigma_S}$ is the jump across the shock at the shock foot). $\nabla_{tg}$ is the tangent gradient (the covariant derivative on the surface) and $\hat{n}_{\sigma_S}$ is the normal vector to the shock boundary curve $\sigma_S$ but is otherwise tangent to the surface.

In order to provide further testing to the approaches described in this paper, we now compute a three-dimensional transonic case, the ONERA M6 wing at $\alpha = 3.06°$, sideslip angle $\beta = 0°$ and $M_\infty = 0.84$. Surface pressure contours are shown in Fig. 15 where the characteristic lambda shock structure is visible.

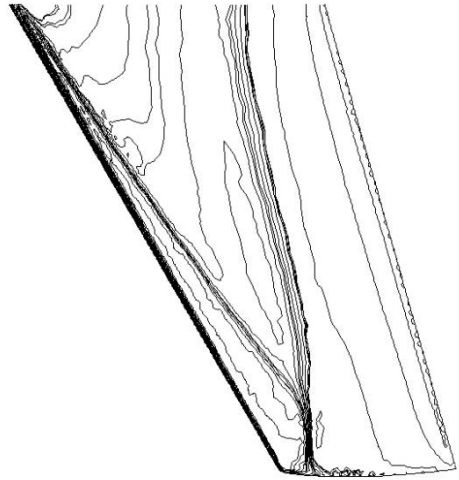

**Fig. 15. ONERA M6. Surface pressure contours**

Mesh adaptation is carried out again with the approaches described above, with the only difference that the percentage of new points is set to 80% in adaptive refinement.

The initial mesh has $\sim 41\times 10^3$ nodes and $\sim 205\times 10^3$ tetrahedral elements (Fig. 16).

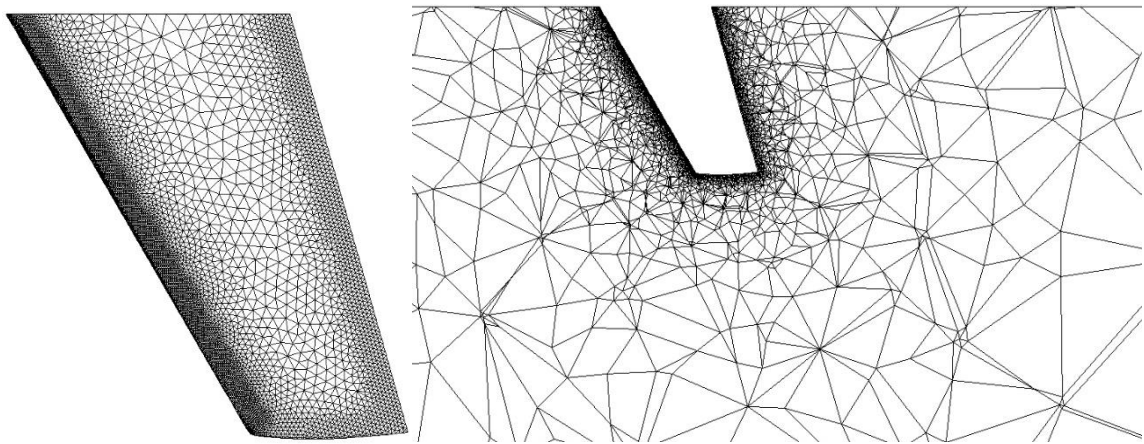

**Fig. 16. ONERA M6. Initial mesh. Left: Close-up view of the unstructured mesh on the top surface of the wing. Right: cut through the wing's horizontal symmetry plane.**

Convergence results for drag and lift are shown in Fig. 17. Errors are established with respect to reference values obtained using Richardson extrapolation from computations in a sequence of globally refined meshes. For the drag coefficient error convergence, both entropy-based approaches show a similar trend: while both show error convergence, they perform worse than uniform refinement at every stage and even worse than feature based adaptation at the earliest stages. This is likely due to the over refinement of the shock and wing-tip vortex regions that is clearly noticeable in the adapted meshes (Fig. 18). On the other hand, the combined entropy adjoint $\times$ drag adjoint, as well as the Oswatitsch adjoint, give results comparable to the (uncorrected) drag-adjoint based ones at the cost of an increased computational effort relative to the pure entropy-based approaches. Accordingly, the corresponding adapted meshes show a more even refinement pattern that is closer to the specific drag-based adaptation. The Oswatitsch case is particularly significant, as in 3D the Oswatitsch formula does not account for induced drag.

At any rate, it is puzzling to see that only the drag adjoint-corrected strategy clearly outperforms uniform refinement for this case. This is likely due to the particular adaptation strategy chosen. The original publication [33] does not show a comparison with uniform refinement in the 3D case, but error convergence trends are similar to the ones shown here. Dwight does note some unusual features such as the fact that the adjoint-corrected values flatten out and only decrease at a very low rate. This is likely due to the fact that the adjoint correction is very effective in eliminating the part of the error due to dissipation, but not so the remaining error, which eventually dominates as the adaptation progresses.

Lift convergence results are much less conclusive, as there are substantial oscillations and even the lift-adjoint corrected values perform worse than uniform refinement (but note that all values shown are within 0.5 lift counts of the reference value). The noisy convergence patterns makes it difficult to identify meaningful trends, but it is perharps noticeable that adaptation with the entropy adjoint indicator gives fairly good results, especially at the earliest stages.

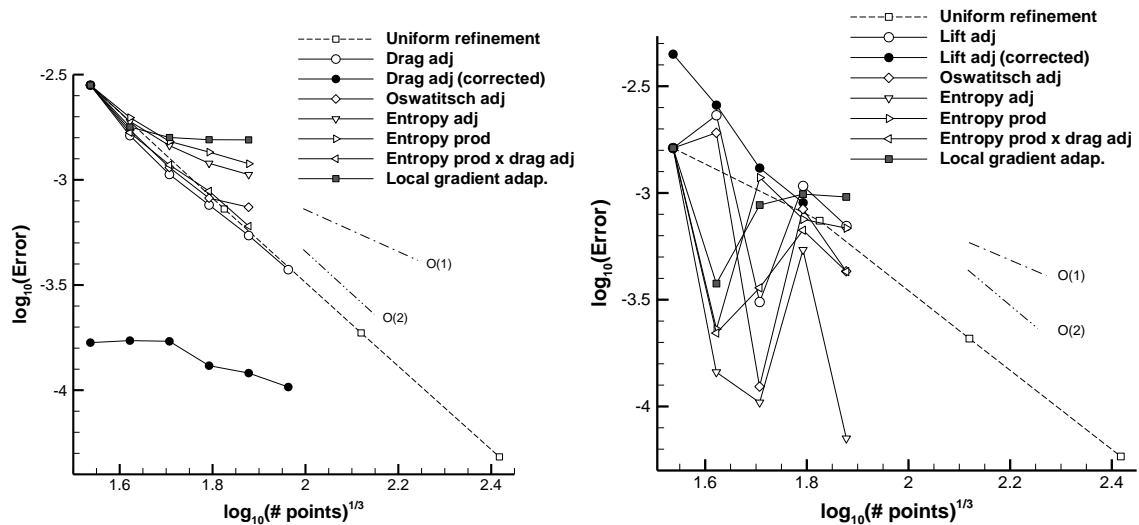

**Fig. 17. ONERA M6. Drag (left) and lift (right) convergence**

**Drag-adjoint**

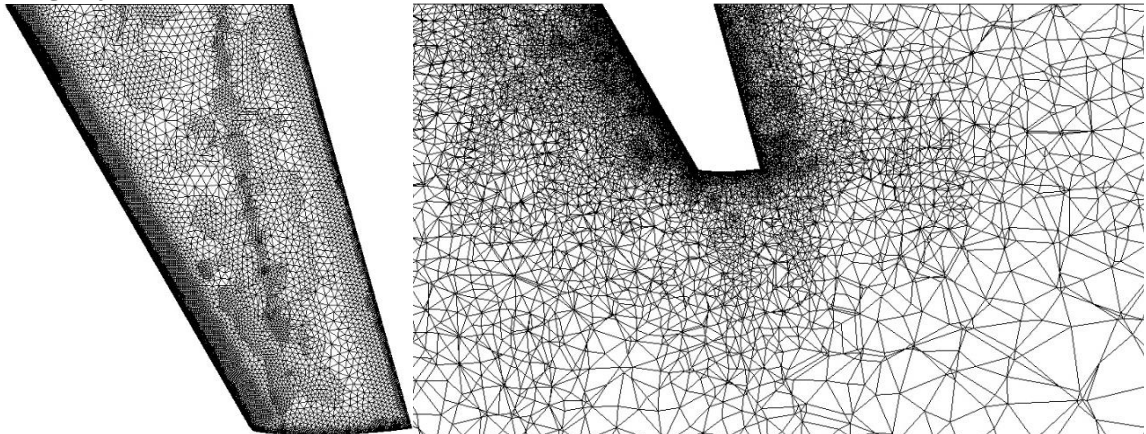

**Lift adjoint**

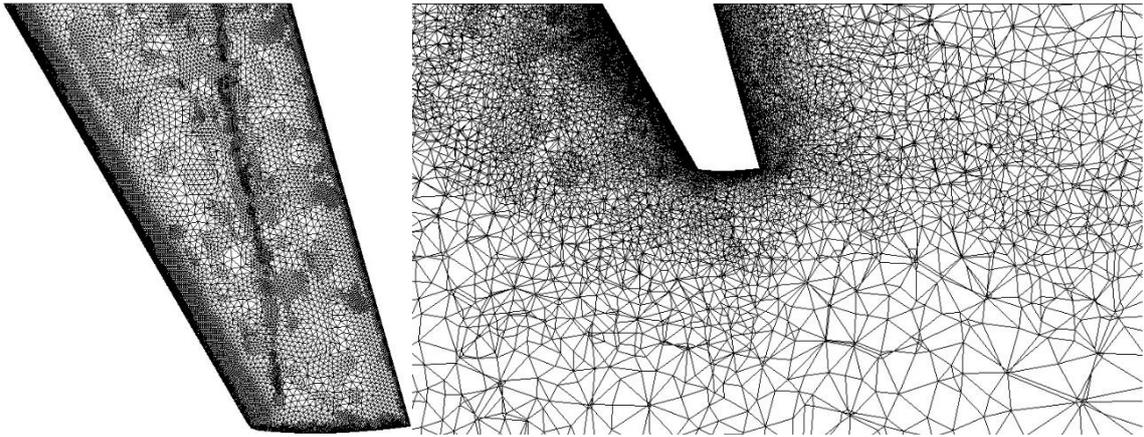

**Entropy adjoint**

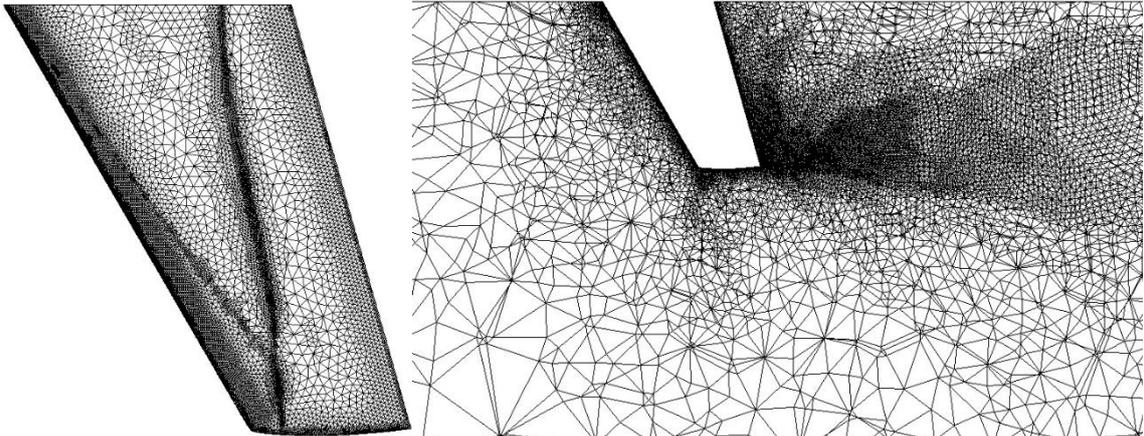

**Entropy production**

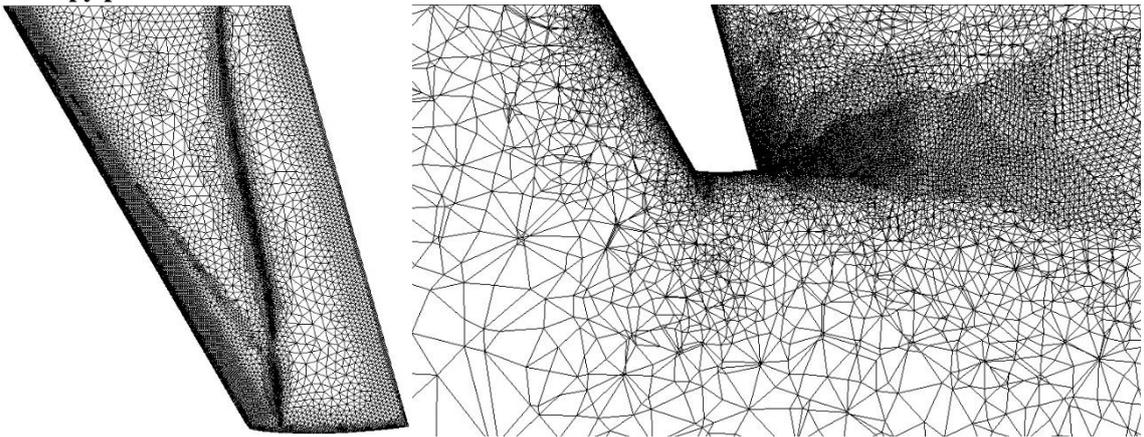

**Oswatitsch adjoint**

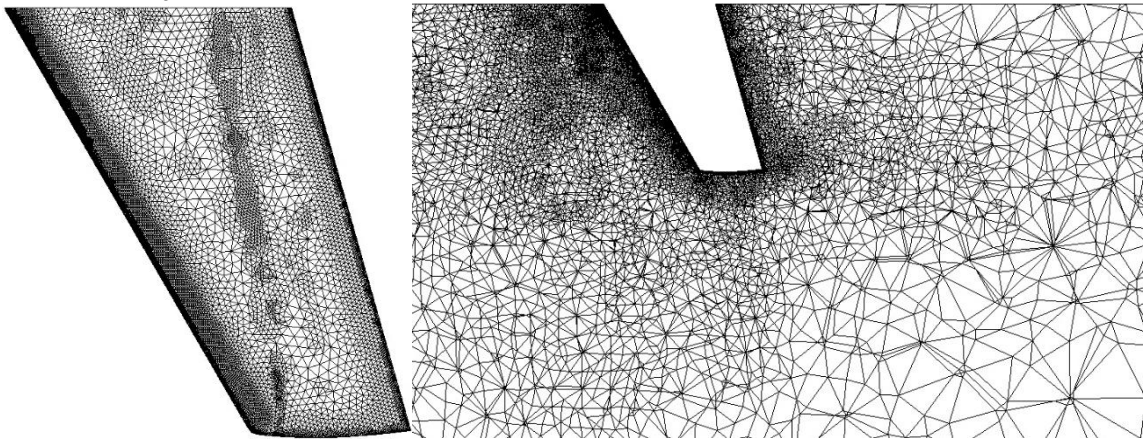

**Entropy production × drag-adjoint sensor**

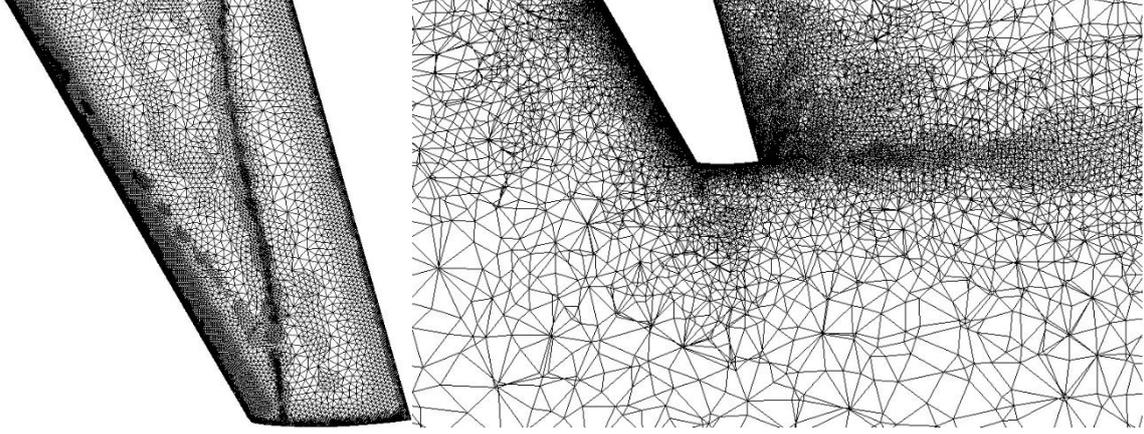

**Fig. 18. ONERA M6. Adapted meshes after 4 refinement cycles.**

## 6  2D Navier-Stokes equations

### 6.1  2D laminar flow

The entropy function (41) is unique in that the entropy variables (42) also symmetrize the viscous terms in the Navier-Stokes equations [16]. If the (steady) Navier-Stokes equations are written as

$$R^{NS}(U) = \nabla \cdot (\vec{F} - \vec{F}^v) = \nabla \cdot \vec{F} - \nabla \cdot (K \cdot \nabla U) = 0 \qquad (65)$$

where $\vec{F}^v = K \cdot \nabla U = (0, \sigma_{i1}, \sigma_{i2}, \sigma_{i3}, u_j \sigma_{ij} + k\partial_i T)^T$ are the viscous fluxes, with $\sigma_{ij}$ the viscous stresses and $k$ the coefficient of thermal conductivity, the entropy variables obey the equation

$$\vec{A}^T \cdot \nabla v - U_v^{-1} \nabla_i (U_v K_{ji}^T \nabla_j v) = \vec{A}^T \cdot \nabla v - U_v^{-1}(\nabla_i U_v) K_{ji}^T \nabla_j v - \nabla_i (K_{ji}^T \nabla_j v) = 0 \qquad (66)$$

which follows from (65) by a simple change of variables $U \to v$ ( $\vec{A}^T \cdot \nabla v - U_v^{-1} \nabla_i (U_v K_{ji}^T \nabla_j v) = U_v^{-1} \nabla \cdot (\vec{F} - \vec{F}^v)$ ). As pointed out in [16], (66) is not a viscous adjoint equation (see eq. (71) below for comparison) and hence the entropy variables are no longer adjoint variables, but they can still be used to "target" spurious entropy production for error estimation and control purposes as follows. Multiplying Eq. (65) by $v^T$ yields the following viscous conservation equation

$$v^T R^{NS}(U) = \nabla \cdot \vec{\Phi} - v^T \nabla \cdot (K \nabla \cdot U) = 0 \qquad (67)$$

Now since $v^T \nabla \cdot \vec{F}^v = -(\sigma : \nabla u + \nabla \cdot (k \nabla T))/(RT)$, where $\sigma : \nabla u = \sigma_{ij} \partial_i u_j$, Eq. (67) is in fact the viscous entropy balance equation [13]

$$\nabla \cdot (\rho \vec{u} s) = \frac{1}{T} \sigma : \nabla u + \frac{1}{T} \nabla \cdot (k \nabla T) \qquad (68)$$

Hence, linear perturbations to the following objective function

$$J = -\frac{1}{R}\int_\Omega \left( \nabla \cdot (\rho \vec{u} s) - \frac{1}{T}(\sigma : \nabla u + \nabla \cdot (k \nabla T)) \right) d\Omega$$
$$= \int_{\partial\Omega} \vec{\Phi} \cdot \hat{n} dS + \int_\Omega \frac{1}{RT}(\sigma : \nabla u + \nabla \cdot (k \nabla T)) d\Omega = \int_\Omega v^T R^{NS}(U) d\Omega \tag{69}$$

measuring spurious entropy production, can be computed as weighted residual perturbations

$$\delta J = \int_\Omega v^T \delta R^{NS}(U) d\Omega \tag{70}$$

In order to target the far-field entropy flux directly, there are two possible avenues. The first possibility, explained in [17], uses the entropy variables and involves two inviscid residual evaluations, one with the computed solution and one with a suitable approximation of the exact solution, and will not be pursued here. The second possibility involves the Oswatitsch adjoint. As above, it is possible to derive the adjoint scheme in this case. Adjoint analysis for viscous flows are quite standard now (see for example [47]) and we will not go through the derivation here. We will just state the basic results. The cost function is again $J^{OSW} = \int_{S^\infty} \vec{\Phi} \cdot \hat{n} dS$ and the associated adjoint state $\psi = (\psi_0, \psi_1, \psi_2, \psi_3)^T$ obeys the adjoint equation

$$\vec{F}_U^T \cdot \nabla \psi - \nabla U^T \cdot \frac{\partial K^T}{\partial U} \cdot \nabla \psi + \nabla \cdot (K^T \cdot \nabla \psi) = 0 \tag{71}$$

subject to the boundary conditions

$$(v - \psi)^T (\vec{F}_U \cdot \hat{n}) \delta U \Big|_{S_\infty} = 0,$$
$$\vec{\varphi}\Big|_{S_w} = (\psi_1, \psi_2)\Big|_{S_w} = 0, \tag{72}$$
$$\partial_n \psi_3 \Big|_{S_w} = 0$$

(adiabatic boundary conditions $\partial_n T\big|_{S_w} = 0$ have been assumed for the primal flow at the wall). The generalization to 3D is immediate.

### 6.2 Numerical tests

We now examine viscous flow past a NACA 0012 airfoil with $M_\infty = 0.5$, $\alpha = 2°$ and Reynolds number $Re = 5000$. The initial mesh, which is shown in Fig. 19, is a hybrid, unstructured mesh with about 30 structured layers of quadrilaterals in the boundary layer. The far-field boundary is approximately 100 chord-lengths away from the airfoil.

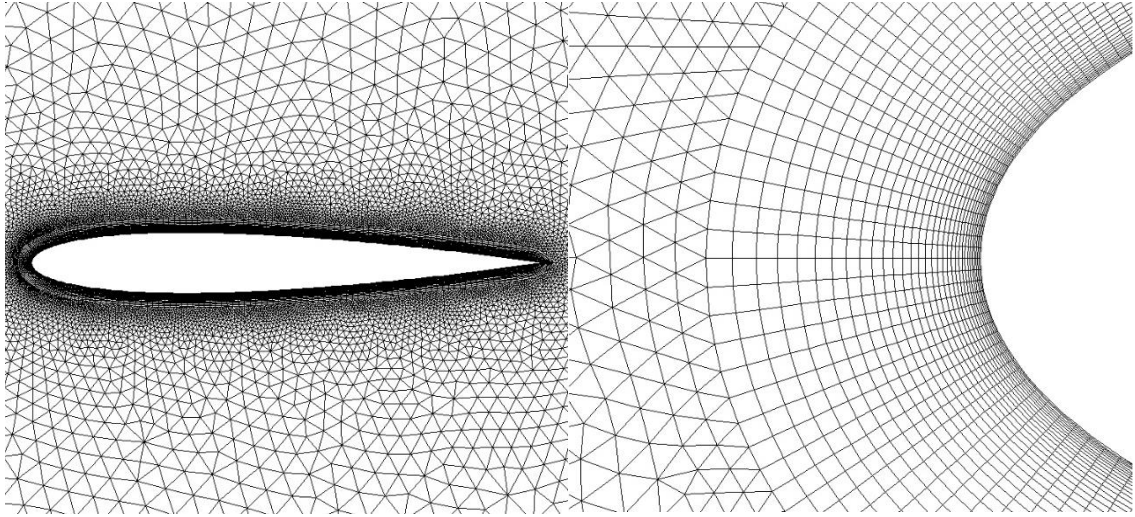

**Fig. 19.** NACA0012 $M_\infty = 0.5$, *Re* = 5000 and $\alpha = 2°$. Initial mesh.

Mesh adaptation is again performed on the above mesh with different strategies including near-field drag and lift, entropy adjoint and Oswatitsch adjoint-based sensors as well as adaptation based on the entropy production of the spatial scheme

$$\Pi_{ij} = \frac{1}{2}(\Delta v_{ij}^T(\vec{n}_{ij} \cdot \vec{f}_{ij}^* + d_{ij} - \vec{n}_{ij} \cdot \vec{f}_{ij}^{(v)}) - \Delta\vec{\Theta}_{ij}) \tag{73}$$

where $\vec{f}_{ij}^{(v)}$ are the viscous fluxes[3]. For the adjoint-based strategies, Dwight's indicator is used. Adaptation in the triangular region proceeds as explained above, while adaptation in the boundary layer proceeds by edge bisection in the streamwise (i.e., parallel to the airfoil surface) direction only.

Fig. 20 shows the convergence of the near-field drag and lift coefficients for the different strategies.

Drag allows for a cleaner comparison, and it shows that the entropy-based strategies yield nearly indistinguishable results, both slightly outperforming the (uncorrected) drag and Oswatitsch adjoints, which also perform comparably. Notice that all output-adapted cases converge at less than first-order after the third adaptation cycle. This reduced rate is likely due to the semifrozen adaptation approach adopted here, whereby the mesh density along the wall-normal direction remains frozen, which has been observed to interfere with the adjoint-based adaptation procedure [48].

Lift convergence results, on the other hand, wander considerably (even for uniform refinement, so the reference value is taken from a globally refined mesh with about 15 million points). On the average, the entropy adjoint based adaptation outperforms all other strategies except the output-based adaptation (results with the Oswatitsch adjoint are better at the first adaptation cycle, but they tend to stall at intermediate stages).

---

[3] In Tau, the viscous flux of a quantity $\phi$ under the effect of viscosity $\mu$ across the dual grid face associated with edge *ij* is given by $\mu \nabla \phi_{ij} \cdot \vec{n}_{ij}$, where $\nabla \phi_{ij} = \frac{1}{2}(\nabla \phi_i + \nabla \phi_j)$ in the plane perpendicular to $\Delta \vec{x}_{ij} = \vec{x}_i - \vec{x}_j$ and $\nabla \phi_{ij} = (\phi_i - \phi_j)\frac{\Delta \vec{x}_{ij}}{\Delta \vec{x}_{ij} \cdot \Delta \vec{x}_{ij}}$ in the direction of $\Delta \vec{x}_{ij}$. Here, $\nabla \phi_i$ and $\nabla \phi_j$ are the gradients of $\phi$ in the dual grid cells *i* and *j* obtained by either Green-Gauss or least-squares reconstruction.

The adapted meshes after 4 adaptation iterations are shown in Fig. 21. All three adjoint-based approaches target strongly the region close to the airfoil, including the wake and the incoming stagnation streamline, more so for the Oswatitsch and near-field drag adjoint indicators. The Oswatitsch adjoint also targets intensely the entire wake all the way to the far-field, a feature that is shared by both entropy-based approaches.

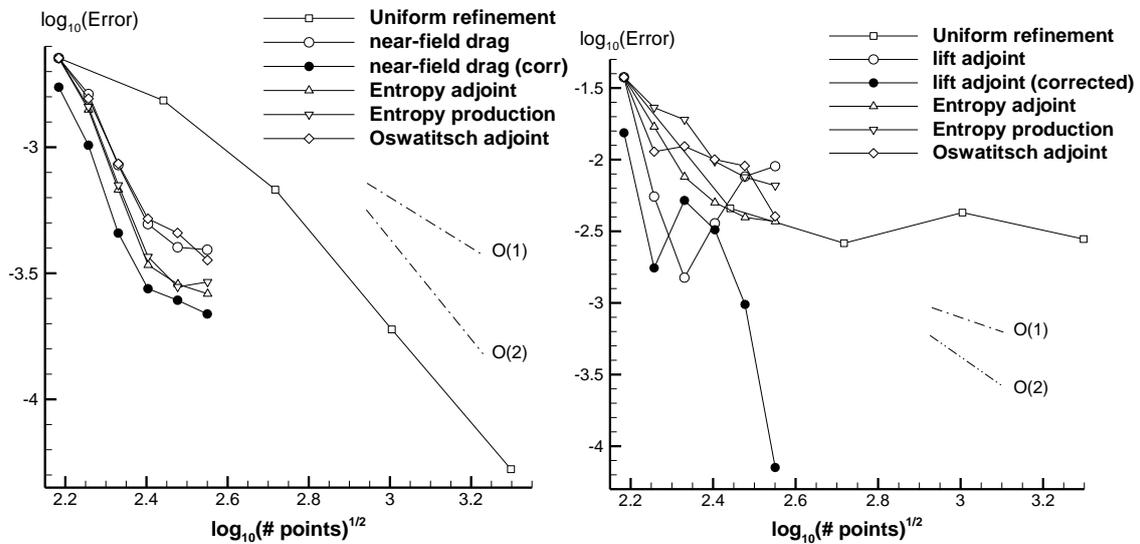

**Fig. 20.** NACA0012 $M_\infty = 0.5$, $Re = 5000$ and $\alpha = 2°$. **Convergence of near-field drag for different adaptation schemes.**

**Drag adjoint**

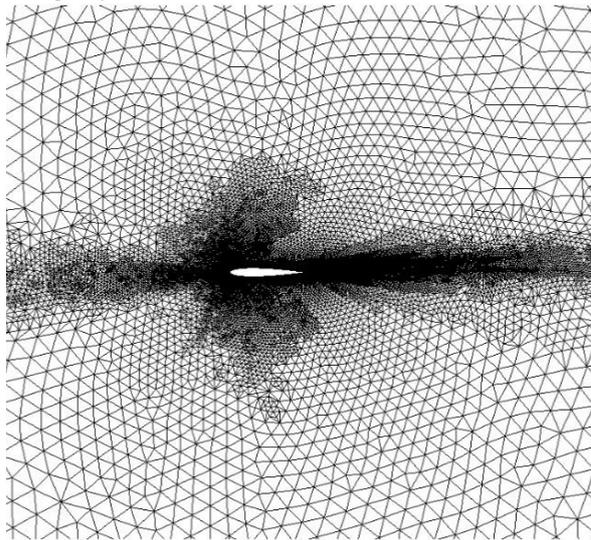
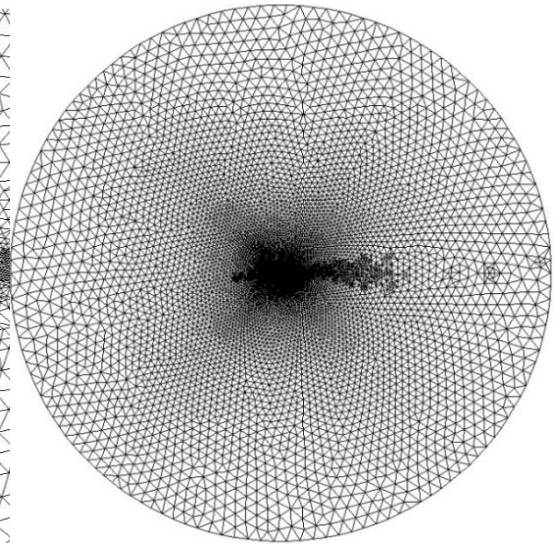

**Lift adjoint**

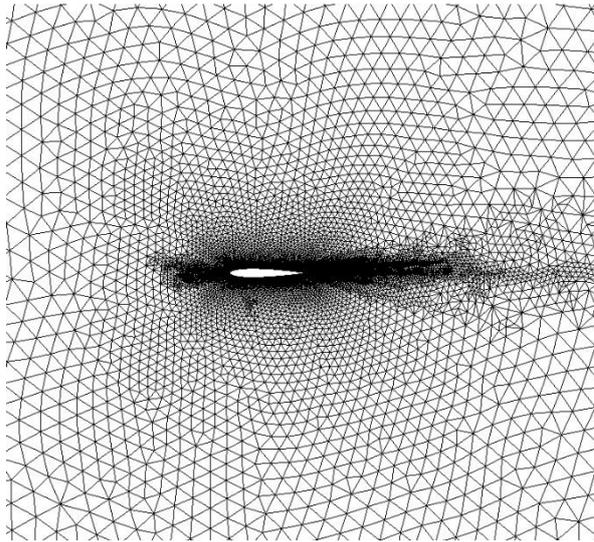
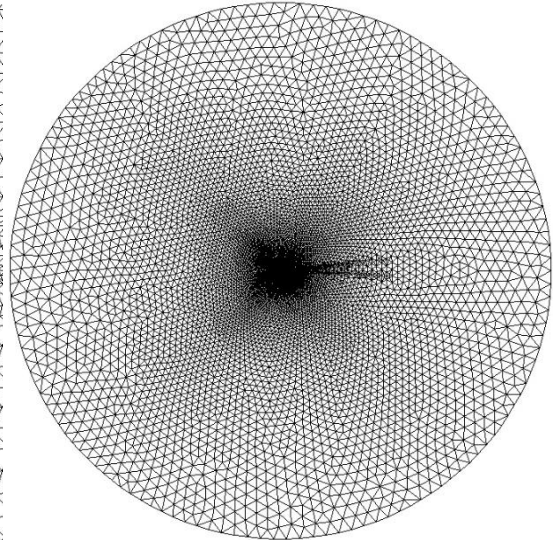

**Oswatitsch adjoint**

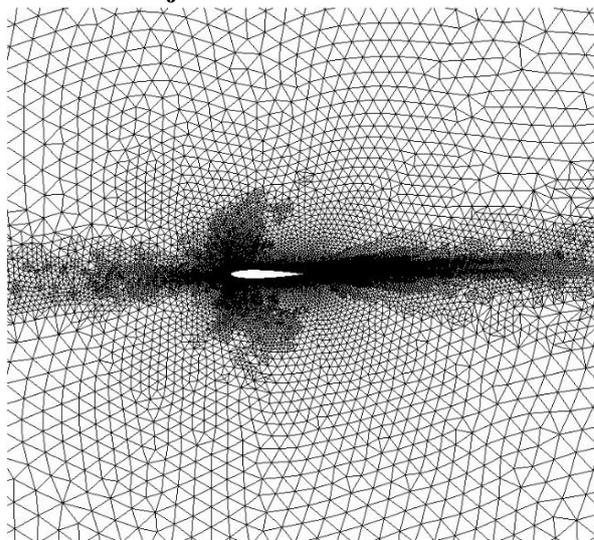
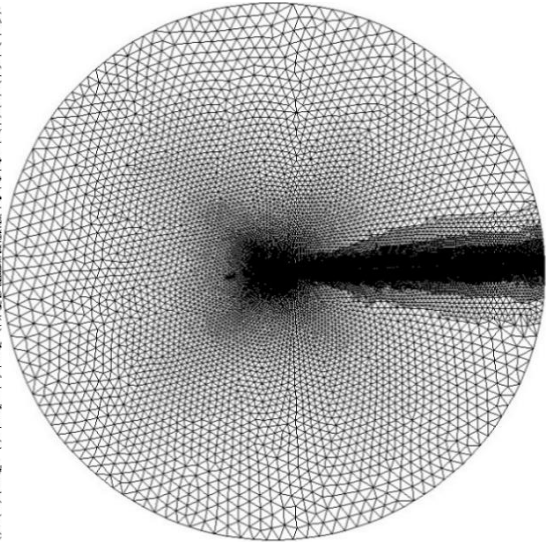

**Entropy adjoint**

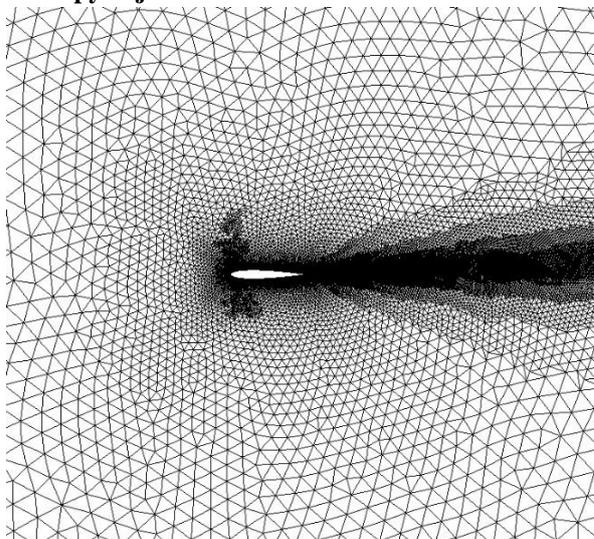
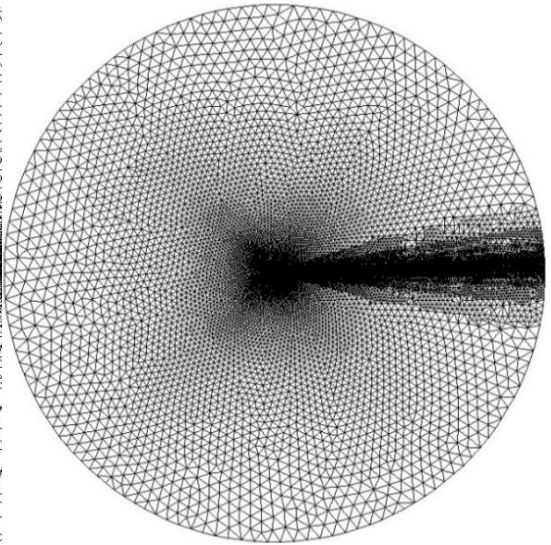

**Entropy production**

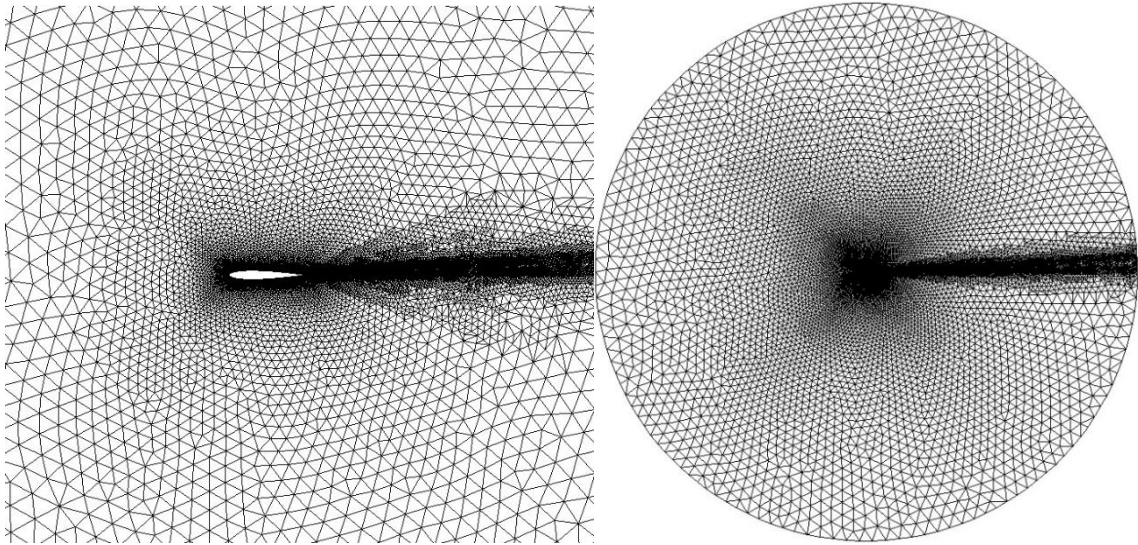

**Fig. 21.** NACA0012 $M_\infty = 0.5$, *Re* = 5000 and $\alpha = 2°$. Adapted meshes after 4 adaptation cycles.

## 7 Discussion

The present paper has focused on two main issues: the relation of entropy variables (or "entropy adjoint") and the adjoint solution based on the far-field entropy flux (that we have called the "Oswatitsch adjoint"), and the comparative performance of mesh adaptation procedures built on the above solutions. We will now briefly summarize the findings concerning each topic.

The Oswatitsch adjoint is dual by construction to the far-field entropy flux, and is thus directly related to aerodynamic drag. The price to pay is, of course, that a separate adjoint solution needs to be computed. The entropy variables, on the other hand, are related (adjoint in the inviscid case, and "quasi"-adjoint in the viscous –laminar– case) to the integrated entropy residual, which can be expressed as an entropy balance comprising the entropy entering and leaving the domain and the entropy generated within the domain, including entropy created at shocks and boundary layers. Hence, the Oswatitsch adjoint is related to physical entropy production, while the entropy adjoint is related to spurious entropy generation. Both approaches intersect for isentropic cases, where they are identical. Likewise, in viscous cases the entropy adjoint can also be used to target drag using the weighted inviscid residual term as explained in [17].

The above distinctions are somehow blurred at the numerical level. As the flow residuals weighted with the Oswatitsch adjoint target errors in the entropy flux, they clearly include spurious entropy generation. On the other hand, the residuals weighted with the entropy variables target regions of spurious entropy production (which are errors in the entropy residual) but also regions (such as shocks) of physical entropy production.

The next question is the practical application of these approaches. We have chosen here to focus on mesh adaptation applications. Both approaches have been tested in shocked inviscid quasi-1D, 2D and 3D flows, as well as in 2D viscous flows, and compared with conventional output-based adjoint mesh adaptation, as well as with adaptation based on the exact numerical entropy production of the spatial discretization.

The Oswatitsch adjoint tends to mimic the drag convergence behavior of the near-field drag adjoint-based adaptation, even in 3D where the Oswatitsch formula does not account for induced drag. The adapted meshes are likewise similar, with the Oswatitsch adjoint targeting the wake more intensely than the near-field drag adjoint but avoiding the over-refinement of the shock regions that is present in both entropy-based approaches. The positive result in 2D cases may have been expected given that, as explained above, the Oswatitsch adjoint also targets regions of spurious entropy production which, as pointed out in [17], accounts for the difference between near-field and Oswatitsch drags.

The entropy adjoint and entropy production approaches, on the other hand, have the advantage of not requiring a separate adjoint solution, but have been seen to perform worse than the output adjoints in inviscid shocked cases. Likely, the creation of entropy at the shocks is leading to over-refinement in those regions, which may be counter-productive as far as prediction of output quantities of engineering interest is concerned. This tendency can be alleviated by weighting the sensor with cell-size dependent factors, but a more physically-based solution is worth investigating. As it turns out, correcting the entropy production sensor to eliminate physical sources of entropy production is relatively easy in the quasi-1D case, but the results are identical to the uncorrected sensor. Such corrections are very hard (if at all possible) to apply in higher dimensions, so alternative routes are explored, such as combining the entropy production sensor with an output-based adjoint sensor either by masking (following a suggestion by Doetsch and Fidkowski), simple addition or multiplication, resulting in improved predictions which however incur in additional computational costs. Alternating entropy-production-based and output-based adaptation has not been found to lead to any significant improvement.

The situation is different in viscous cases, at least in the subsonic, laminar case that we have analyzed, where entropy-based approaches have been seen to outperform output-based approaches in terms of drag error convergence. It remains to be seen the effect of including turbulence or viscous shocks.

Overall, the entropy adjoint/entropy production approach has the obvious advantage of not requiring a separate adjoint solution, which can be time consuming and somewhat tricky, for robustness reasons, since adjoint solutions become increasingly difficult to converge as the adaptation progresses; on the other hand, results with the entropy adjoint are disappointing for shocked flows, but much more promising for viscous flows even with one single residual evaluation.

The Oswatitsch adjoint, on the other hand, yields results very similar to the near-field drag adjoint (recall that we are measuring the performance of the Oswatitsch adjoint, which is associated with a far-field drag measurement, with respect to a near-field drag computation).

We end with two considerations which concern the relevance of the Oswatitsch adjoint approach and the influence of the particular adaptation strategy adopted. Regarding the first issue, and as explained in the introduction, the far-field evaluation of the entropy flux by itself (and not just as a surrogate for aerodynamic drag) is relevant in turbomachinery flows as a means to compute loss, which has sources other than aerodynamic drag. The adjoint formulation presented here could be of relevance in the numerical estimation of loss for these flows, as well as for optimal shape design.

Finally, the particular adaptation strategy chosen here may of course be debated, but the idea here was to compare all the approaches within a single, simple, adaptation setting. The adopted adaptive strategy focuses on the elimination of artificial dissipation errors, and this should be kept in mind when assessing and extrapolating these results. However, artificial viscosity contributes significantly to spurious entropy generation, especially in inviscid flows [5], which supports the choice of adaptation strategy. Besides, the overall conclusions are not significantly different from what can be found in the published literature, giving more confidence in the generalizability of the results.

## Acknowledgements

This work has been supported by INTA and the Spanish Ministry of Defence under the research program "Termofluidodinámica" (IGB99001). The 2D and 3D computations were carried out with the TAU Code, developed at DLR's Institute of Aerodynamics and Flow Technology at Göttingen and Braunschweig, which is licensed to INTA through a research and development cooperation agreement.